\documentclass[longauth, traditabstract]{aa}
\usepackage{txfonts}
\usepackage{graphicx}
\usepackage{natbib}
\usepackage{xspace}
\usepackage{color}
\bibpunct{(}{)}{;}{a}{}{,}

\def\mpcoh{\,h^{-1}{\rm Mpc}}
\def\hompc{\,h\,{\rm Mpc}^{-1}}
\def\gpcoh{\,h^{-1}{\rm Gpc}}
\def\msunoh{\,h^{-1}{\rm M}_\odot}
\def\kmsmpc{\,{\rm km\,s^{-1}Mpc^{-1}}}
\def\kms{\,{\rm km\,s^{-1}}}

\newcommand{\dif}{{\mathrm d}}

\begin{document}

\title{The VIMOS public extragalactic redshift survey (VIPERS)
    \thanks{based on observations collected at the European Southern
    Observatory, Cerro Paranal, Chile, using the Very Large Telescope
    under programmes 182.A-0886 and partly 070.A-9007.  Also based on
    observations obtained with MegaPrime/MegaCam, a joint project of
    CFHT and CEA/DAPNIA at the Canada-France-Hawaii Telescope (CFHT),
    which is operated by the National Research Council (NRC) of
    Canada, the Institut National des Science de l'Univers of the
    Centre National de la Recherche Scientifique (CNRS) of France, and
    the University of Hawaii.  This work is based in part on data
    products produced at TERAPIX and the Canadian Astronomy Data
    Centre as part of the Canada-France-Hawaii Telescope Legacy
    Survey, a collaborative project of NRC and CNRS.  The VIPERS web
    site is http://www.vipers.inaf.it/. }
}

\subtitle{$\Omega_{\rm m_0}$ from the galaxy clustering ratio measured at $z \sim 1$}
%
%
\author{
 J.~Bel\inst{1,30,3} 
\and  C.~Marinoni\inst{1,2,30}
\and  B.~R.~Granett\inst{3}
\and L.~Guzzo\inst{3,4}
\and  J.~A.~Peacock\inst{5}
\and  E.~Branchini\inst{7,8,9}
\and  O.~Cucciati\inst{10}
\and S.~de la Torre\inst{5}
\and A.~Iovino\inst{3}
\and W.~J.~Percival\inst{26}
\and H.~Steigerwald\inst{1,30}
\and U.~Abbas\inst{11}
\and C.~Adami\inst{12}
\and S.~Arnouts\inst{13,12}
\and M.~Bolzonella\inst{9}           
\and D.~Bottini\inst{10}
\and A.~Cappi\inst{9,14}
\and J.~Coupon\inst{15}        
\and I.~Davidzon\inst{9,16}
\and G.~De Lucia\inst{17}
\and A.~Fritz\inst{10}
\and P.~Franzetti\inst{10}
\and M.~Fumana\inst{10}
\and B.~Garilli\inst{10,13}     
\and O.~Ilbert\inst{12}
\and J.~Krywult\inst{18}
\and V.~Le Brun\inst{12}
\and O.~Le F\`evre\inst{12}
\and D.~Maccagni\inst{10}
\and K.~Ma{\l}ek\inst{19}
\and F.~Marulli\inst{16,20,9}
\and H.~J.~McCracken\inst{21}
\and L.~Paioro\inst{10}
\and M.~Polletta\inst{10}
\and A.~Pollo\inst{22,23}
\and H.~Schlagenhaufer\inst{24,25}
\and M.~Scodeggio\inst{10} 
\and L.~A.~.M.~Tasca\inst{12}
\and R.~Tojeiro\inst{26}
\and D.~Vergani\inst{27,9}
\and A.~Zanichelli\inst{28}
\and A.~Burden\inst{26}
\and C.~Di Porto\inst{9}
\and A.~Marchetti\inst{29,3} 
\and Y.~Mellier\inst{21}
\and L.~Moscardini\inst{16,20,9}
\and R.~C.~Nichol\inst{27}
\and S.~Phleps\inst{24}
\and M.~Wolk\inst{21}
\and G.~Zamorani\inst{9}
}

\offprints{ Bel., J. \\ \email{jbel@cpt.univ-mrs.fr} }

\institute{Aix Marseille Universit\'e, CNRS, CPT, UMR 7332, 13288 Marseille, France  
\and Institut Universitaire de France  
\and INAF - Osservatorio Astronomico di Brera, Via Brera 28, 20122 Milano, via E. Bianchi 46, 23807 Merate, Italy 
\and Dipartimento di Fisica, Universit\`a di Milano-Bicocca, P.zza della Scienza 3, I-20126 Milano, Italy 
\and SUPA, Institute for Astronomy, University of Edinburgh, Royal Observatory, Blackford Hill, Edinburgh EH9 3HJ, UK 
\and Dipartimento di Matematica e Fisica, Universit\`{a} degli Studi Roma Tre, via della Vasca Navale 84, 00146 Roma, Italy 
\and INFN, Sezione di Roma Tre, via della Vasca Navale 84, I-00146 Roma, Italy 
\and INAF - Osservatorio Astronomico di Roma, via Frascati 33, I-00040 Monte Porzio Catone (RM), Italy 
\and INAF - Osservatorio Astronomico di Bologna, via Ranzani 1, I-40127, Bologna, Italy 
\and INAF - Istituto di Astrofisica Spaziale e Fisica Cosmica Milano, via Bassini 15, 20133 Milano, Italy
\and INAF - Osservatorio Astrofisico di Torino, 10025 Pino Torinese, Italy 
\and Aix Marseille Universit\'e, CNRS, LAM (Laboratoire d'Astrophysique de Marseille) UMR 7326, 13388, Marseille, France  
\and Canada-France-Hawaii Telescope, 65--1238 Mamalahoa Highway, Kamuela, HI 96743, USA 
\and Laboratoire Lagrange, UMR7293, Universit\'e de Nice Sophia-Antipolis,  CNRS, Observatoire de la C\^ote d'Azur, 06300 Nice, France 
\and Institute of Astronomy and Astrophysics, Academia Sinica, P.O. Box 23-141, Taipei 10617, Taiwan
\and Dipartimento di Fisica e Astronomia - Universit\`{a} di Bologna, viale Berti Pichat 6/2, I-40127 Bologna, Italy 
\and INAF - Osservatorio Astronomico di Trieste, via G. B. Tiepolo 11, 34143 Trieste, Italy 
\and Institute of Physics, Jan Kochanowski University, ul. Swietokrzyska 15, 25-406 Kielce, Poland 
\and Department of Particle and Astrophysical Science, Nagoya University, Furo-cho, Chikusa-ku, 464-8602 Nagoya, Japan 
\and INFN, Sezione di Bologna, viale Berti Pichat 6/2, I-40127 Bologna, Italy 
\and Institute d'Astrophysique de Paris, UMR7095 CNRS, Universit\'{e} Pierre et Marie Curie, 98 bis Boulevard Arago, 75014 Paris, France 
\and Astronomical Observatory of the Jagiellonian University, Orla 171, 30-001 Cracow, Poland 
\and National Centre for Nuclear Research, ul. Hoza 69, 00-681 Warszawa, Poland 
\and Max-Planck-Institut f\"{u}r Extraterrestrische Physik, D-84571 Garching b. M\"{u}nchen, Germany 
\and Universit\"{a}tssternwarte M\"{u}nchen, Ludwig-Maximillians Universit\"{a}t, Scheinerstr. 1, D-81679 M\"{u}nchen, Germany 
\and Institute of Cosmology and Gravitation, Dennis Sciama Building, University of Portsmouth, Burnaby Road, Portsmouth, PO1 3FX 
\and INAF - Istituto di Astrofisica Spaziale e Fisica Cosmica Bologna, via Gobetti 101, I-40129 Bologna, Italy 
\and INAF - Istituto di Radioastronomia, via Gobetti 101, I-40129, Bologna, Italy 
\and  Universit\`{a} degli Studi di Milano, via G. Celoria 16, 20130 Milano, Italy 
\and Universit\'e de Toulon, CNRS, CPT, UMR 7332, 83957 La Garde, France 
}
%
%
\date{Received --; accepted --}
%

\abstract{
We use a sample of about 22,000 galaxies at $0.65<z<1.2$ from the
VIMOS public extragalactic redshift survey (VIPERS) public data
release 1 (PDR-1) catalogue, to constrain the cosmological model through a measurement
of the galaxy {\it clustering ratio} $\eta_{g,R}$. This statistic has
favourable properties, which is defined as
the ratio of two quantities characterizing the smoothed density field
in spheres of a given radius $R$: the
value of its correlation function on a multiple of this scale, $\xi(nR)$, and its
variance $\sigma^2(R)$.
For sufficiently large values of $R$, this is a
universal number, which captures 2-point clustering information
independently of the 
linear bias and linear redshift-space distortions of the specific galaxy tracers.
In this paper, we discuss how to extend the application of
$\eta_{g,R}$ to quasi-linear scales and how to control and remove
observational selection effects, which are typical of redshift surveys
as VIPERS, in detail.  We verify the accuracy and efficiency of these procedures
using mock catalogues that match the survey 
selection process. 
These results {\bf show} the robustness
of $\eta_{g,R}$ to non-linearities and observational effects, which is
related to its very definition as a ratio of quantities that are similarly affected. 

At an effective redshift $z=0.93$, we measured the value $\eta_{g,R}(15)=0.141\pm 0.013$ at $R=5$ $h^{-1}$ Mpc.  
Within a flat $\Lambda$CDM cosmology and by including the best available
priors on $H_0$, $n_s$ and baryon density, we obtain a matter
density parameter at the current epoch
$\Omega_{m,0}=0.270_{-0.025}^{+0.029}$.  In addition to the great
precision achieved on our estimation of $\Omega_m$ using VIPERS PDR-1,
this result is remarkable because  it appears to be in good agreement
with a recent estimate at $z\simeq 0.3$, which was obtained by 
applying the same technique to the SDSS-LRG catalogue.
It, therefore, supports the robustness of the present analysis.
Moreover, the combination of these two measurements at $z\sim 0.3$ and $z\sim
0.9$ provides us with a very precise estimate of $\Omega_{m,0}=0.274\pm0.017$, which highlights the great consistency between our estimation and other cosmological probes, such as BAOs, CMB, and Supernovae.    
}


\maketitle

\section{Introduction}
The present-day large-scale structure in cosmological matter
distribution is 
formed by the gravitational
amplification of small density perturbations that are a relic of the
early Universe.  The amplitude of these fluctuations as a function of
scale carries unique information about the fundamental cosmological
parameters, which describe the dominant constituents of the Universe
and their densities with the global expansion history.  
Since the primordial density field is commonly thought
to be a random Gaussian process, the Fourier power spectrum, or, its
real-space counterpart, the correlation function provide a
complete statistical description.  Linear growth of small fluctuations
preserves the shape of both, suggesting that
observations of the large-scale structure today should still be able to
extract pristine cosmological information.

This idealized picture is complicated in practice, as
galaxies do not faithfully trace 
the distribution of matter but are undoubtedly {\it biased\/} to some extent.
The reason this is inevitable is because galaxies form within
dark matter haloes, and the more massive haloes form at special
places within the cosmic density field. 
This generates a
large-scale linear bias that gives a two-point correlation function, $\xi(r)$,
of galaxies that is related to that of mass by:
\begin{equation}
\xi_g(r) = b^2(M)\xi_m(r)\quad {\rm if}\quad \xi_m(r) \ll1,
\end{equation}
where the bias parameter, $b$, depends on halo mass  \citep{kaiser84,mowh,scto}.

On small scales where the correlations are non-linear, the correlation
function gains additional contributions, which arise from pairs within
haloes that are massive enough to host more than one galaxy.  Thus, the
galaxy correlation function varies in shape and amplitude for
different classes of galaxies because different classes of galaxy
occupy different masses of haloes to different extents, and the
distribution of galaxies within each halo does not necessary follow
that of the mass \citep[This is the basis of the `halo model' -- see
e.g.][.]{cosh}
Weak non-linearities in the mapping between the
underlying matter density field and the distribution of
collapsed objects also develop on quasi-linear scales \citep{sbd, marinoni},  
as a consequence of the complex interplay between galaxy formation processes and
gravitational dynamics. Additionally,
there may be some scatter in the galaxy properties
found in a set of haloes of the same mass; these effects are
collected under the heading of `stochastic bias' \citep[e.g.][]{dl99, GW07} 
and have been shown to be small but
non-zero in the observed galaxy distribution \citep[e.g.][]{w05}.

The non-linear physics responsible for galaxy formation can be mitigated by surveys covering large volumes,
which can probe the power spectrum closer to the linear regime, or by
focussing on robust features such as Baryonic Acoustic Oscillations
(BAO), the relic of the baryon-photon interactions in the
pre-recombination plasma. These strategies have motivated the large
``low-redshift'' surveys of the past decade (e.g SDSS-3 BOSS
\citealt{eisenstein11,anderson12}; WiggleZ \citealt{drinkwater10,
  blake11}; 2dFGRS \citealt{colless}) and are also a major driver for the next generation of
projects from the ground or space (e.g. BigBoss \citealt{Schlegel};
Euclid \citealt{Laureijs}).

The development of multi-object spectrographs on 8-m class telescopes
during the 1990s triggered a number of deep redshift surveys with
measured distances beyond $z\sim 0.5$ over areas of 1--2 deg$^2$
(e.g. VVDS \citealt{lefevre}, DEEP2 \citealt{newman} and zCOSMOS
\citealt{lilly09}). Even so, it was not until the wide extension of
VVDS was produced \citep{Garilli08} that a survey existed with
sufficient volume to attempt cosmologically meaningful computations at
$z\sim 1$ \citep{Guzzo}.  In general, clustering measurements at
$z\simeq 1$ from these samples remained dominated by cosmic variance,
as dramatically shown by the discrepancy observed between the VVDS and
zCOSMOS correlation functions at $z\simeq 0.8$ \citep{delaTorre10}.
 
The VIMOS Public Extragalactic Redshift Survey (VIPERS) is part of
this global attempt to take cosmological measurements at $z\sim 1$ to
a new level in terms of statistical significance. In contrast to the
BOSS and WiggleZ surveys, which use large-field-of-view ($\sim 1$
deg$^2$) fibre optic positioners to probe huge volumes at low sampling
density, VIPERS exploits the features of VIMOS at the ESO VLT to yield
a dense galaxy sampling over a moderately large field of view ($\sim
0.08$ deg$^2$). It reaches a volume at $0.5<z<1.2$, which is comparable to that
of the 2dFGRS \citep{colless} at $z\sim 0.1$, allowing the
cosmological evolution to be tested with small statistical errors.

In parallel to improving the samples, we can try and devise
statistical estimators that are capable of
probing the density field, while being weakly sensitive to galaxy bias
and non-linear evolution  \citep[e.g.][]{zhang, bmMNRAS}.
This is the motivation for a new statistic introduced by Bel
\& Marinoni (2013; hereafter BM13), who  proposed the use of the
{\it clustering ratio\/} $\eta_{g,R}(r)$, the ratio of the
correlation function to the variance of the galaxy over-density
field $\delta_R$ smoothed on a scale $R$. BM13 argue that this
statistic has a redshift-space form that is independent of the
specific bias of the galaxies used and is essentially only sensitive
to the shape of the non-linear real-space
matter power spectrum in two relatively narrow bandwidths.

BM13 have used numerical simulations to thoroughly discuss 
the level of precision reachable through this method on large linear
scales. In particular, they showed that the clustering ratio measured from the
simulations differs from the theoretical prediction by only $0.1\%$ for some specific choices of
the scales $R$ and $r$ that are involved in the smoothing of the auto-correlation
of the galaxy density field.
They also applied this technique to the SDSS-DR7 Luminous Red Galaxy
sample with $z<0.4$ by obtaining an estimate of the present matter
density ($\Omega_{m,0}=0.283\pm 0.023$), which is $20\%$ more precise 
than the measurement of \citet{per10} from a BAO analysis of the same sample. This demonstrates the
sensitivity of relative  clustering measurements made on
different scales to $\Omega_{m,0}$, which controls the horizon scale
at matter-radiation equality and the corresponding turn-over in the
growth rate of fluctuations.
 
The $\eta$ statistics remains sensitive to non-linearities if we probe
small enough scales. Since the statistical errors decline as we add
small-scale data, there is a careful balance to be struck in attaining
a result that has a small error bar, while lacking significant
systematic errors. Non-linearities in themselves are unimportant, since
the non-linear evolution of the matter power spectrum is in effect a
solved problem at the level of precision we need and can be predicted
for any model we wish to test. 
The non-trivial problem is eventually inadequate for the local bias
hypothesis \citep{FG} - i.e., a possible dependence of the bias parameter  on the separation
scale $r$ (cfr. eq. (1)).  We address this issue carefully in the present
paper, using realistic simulated galaxy distributions, and show that
the likely degree of scale dependence of the biasing relation in Fourier space 
is small enough that it does not constitute a significant uncertainty in the interpretation of our
measurements of $\eta$.

In this paper, we measure the $\eta_{g,R}$ statistic from the VIPERS
public data release 1 (PDR-1) redshift catalogue, by including $\sim 64\%$ of the final number of
redshifts expected at completion  (see \citealt{Guzzo13}, hereafter Paper I, for a detailed
description of the survey data set).  The paper is organized as follows. In \S 2, we introduce the VIPERS
survey and the features of the PDR-1 sample. In \S 3, we review the basics of the
$\eta$-test,  and we discuss how the galaxy clustering ratio
is estimated from VIPERS data. In \S 4, we present the formalism that
allows us to also predict the amplitude of this new cosmological observable in the non-linear regime, and we  assess its 
overall viability  with numerical simulations.  This analysis
helps us to identify the range of the parameter space  
where cosmological results can be meaningfully interpreted.   In \S 5,
we discuss the correction schemes adopted to account for the
imprint of the specific VIPERS mask and selection effects, testing
them with VIPERS mock surveys. Cosmological results are presented in \S 6 and
conclusions are drawn in \S 7. 

Throughout this paper, the Hubble constant is parameterized via $h=H_0/100\kmsmpc$, 
and all magnitudes in this paper are in the AB system \citep{og}.
We do not give an explicit AB suffix.  We do not fix cosmological
parameters to a fiducial set of value. For consistency, we reconstruct the galaxy clustering ratio and analyse data in
any tested cosmology. 

\section{Data}

\begin{figure*}
\centerline{\includegraphics[width=180mm,angle=0]{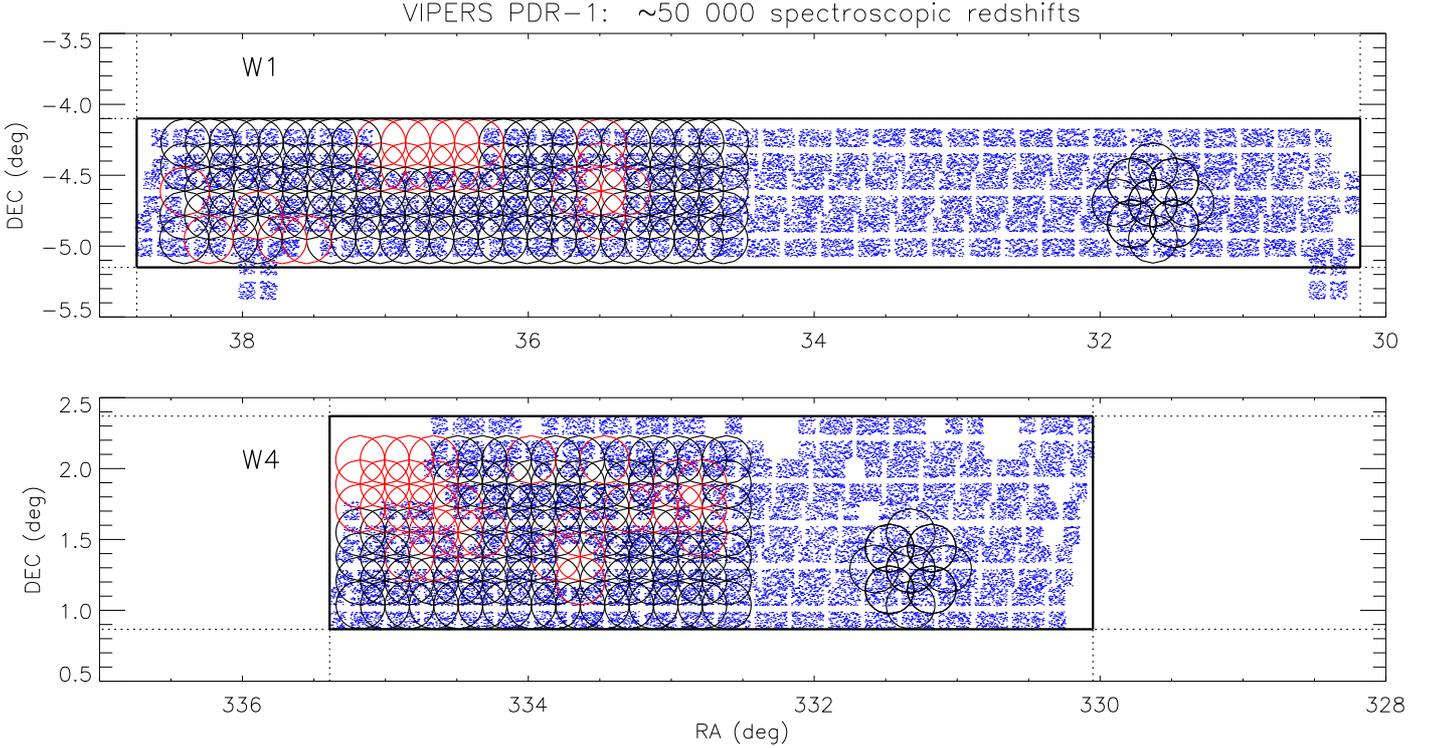}}
\caption{\small  Distribution on the sky of the galaxies with measured
  redshifts in the VIPERS PDR-1 catalogue within the W1 and W4
  fields (blue points).  The  pattern described by
  the gaps separating the VIMOS quadrants, as well as the areas not covered
  by the PDR-1, are clearly visible.  Superimposed to
  the spectroscopic data, we also plot  for illustrative purposes the projection of a subset
  of the spherical cells used to estimate the variance
  $\sigma_{g,R}^2$ (regular grid on the left) with the typical motif used to
  estimate the $2$-point correlation function of the smoothed
  density field $\xi_{g,R}(r)$ (right). The spheres have a radius of
  $R=5\mpcoh$, and the chosen set is placed
  at redshift $z=0.66$.  Note that the distance between the centres of
  $2$ opposite spheres in the motif is exactly equal to the chosen
  correlation scale $r$ (we show here the case $r=3R$). Red circles
  correspond to spheres that have been rejected in the
  clustering analysis because they are heavily affected by the survey mask
  (See discussion in \S~\ref{sec:angular}.) }  
\label{w1w4}
\end{figure*}
 As already mentioned in this paper, we use the new data of the VIPERS, which is being built using the
VIMOS spectrograph at the ESO VLT. The survey target sample is
selected from the Canada-France-Hawaii Telescope Legacy Survey Wide
(CFHTLS-Wide) optical photometric catalogues \citep{CFHTLS}. The VIPERS
covers $\sim24$ deg$^2$ on the sky divided over two areas within the
W1 and W4 CFHTLS fields. Galaxies are selected to a limit of
$i_{AB}<22.5$, which further applies a simple and robust $gri$ colour
pre-selection, as to effectively remove galaxies at $z<0.5$. Coupled
to an aggressive observing strategy (Scodeggio et al. 2009), this
allows us to double the galaxy sampling rate in the redshift range of
interest with respect to a pure magnitude-limited sample ($\sim
40\%$). At the same time, the area and depth of the survey result in a
fairly large volume, $\sim 5 \times 10^{7}$ h$^{-3}$ Mpc$^{3}$,
analogous to that of the 2dFGRS at $z\sim0.1$. Such combination of
sampling and depth is quite unique over current redshift surveys at
$z>0.5$. The VIPERS spectra are collected with the VIMOS multi-object
spectrograph \citep{vimos_ref} at moderate resolution ($R=210$), using
the LR Red grism, providing a wavelength coverage of
5500-9500${\rm\AA}$ and a typical redshift error of $141(1+z)$ km
sec$^{-1}$ . The full VIPERS area of $\sim 24$ deg$^2$ is covered
through a mosaic of 288 VIMOS pointings (192 in the W1 area and 96 in
the W4 area). A discussion of the survey data reduction and management
infrastructure is presented in \cite{Garilli2012}. An early subset of
the spectra used here is analysed and classified through a Principal
Component Analysis (PCA) in \cite{Marchetti2012}.

A quality flag is assigned to each measured redshift based on the
quality of the corresponding spectrum. Here and in all parallel VIPERS
science analyses, we use only galaxies with flags 2 to 9 inclusive,
which correspond to a global redshift confidence level of 98\%. The
redshift confirmation rate and redshift accuracy have been estimated
using repeated spectroscopic observations in the VIPERS fields. A more complete description of the survey
construction from the definition of the target sample to the actual
spectra and redshift measurements is given in the parallel survey
description paper (Paper I).

The data set used in this and the other papers is the VIPERS PDR-1 catalogue, which
was made publicly available in September 2013.  This includes
$55,359$ objects which are spread over a global area of $8.6\times 1.0$ deg$^2$
and $5.3\times 1.5$ deg$^2$, respectively, in W1 and W4. It corresponds
to the data frozen in the VIPERS database at the end of the 2011/2012
observing campaign with 64\% of the final expected survey yield. 
For the specific analysis presented here, the sample has been
further limited to its higher-redshift part, selecting only galaxies
with $0.65<z< 1.2$. The reason for this selection is related to the necessity of
  covering a minimum physical size in the declination direction,
  which, given the angular aperture of the survey, cannot be obtained
  for smaller distances (more details are given at the end of  \S 3.2).
 This reduces the usable sample to 13,688 and
12,923 galaxies in W1 and W4, respectively (always with quality flag $\ge 2$,
as defined earlier).  The corresponding volumes of the two samples are 1.3
and 1.2 $\times 10^{7}$ h$^{-3}$ Mpc$^{3}$ with a spanned largest linear
dimension at $z=1.2$ of $\sim 400$ and $250\mpcoh$, respectively.

\section{Method}

Here we briefly review the basic features of the {\sl clustering
ratio} cosmological test as proposed by
BM13. We then discuss the procedure adopted to estimate it from a
redshift survey like VIPERS.
 
\subsection{The clustering ratio}
Let $\delta_{g,R}({\bf x})$ be the galaxy overdensity field that is smoothed on a
scale $R$ at the real-space position ${\bf x}$ (In the
following, we assume that smoothing is done by convolving the field
with a spherical Top Hat window of radius $R$.)  Its variance
$\sigma_{g,R}^2= \langle \delta_{g,R}^2({\bf x}) \rangle_c$  and 
$2$-point correlation function $\xi_{g,R}(r)= \langle
\delta_{g,R}({\bf x})\delta_{g,R}({\bf x} +{\bf r})\rangle_c$ 
(here $\langle...\rangle_c$ indicates cumulant expectation values)  
can be combined to define the  galaxy clustering ratio  
$\eta_{g,R}(r,{\bf p})\equiv \frac{\xi_{g,R}(r)}{\sigma_{g,R}^2}$ ($p$ emphasis the dependency regarding to cosmological parameters). 
This statistic is a measure of the
`slope'  of the correlation function,  which is smoothed over a particular  double kernel 
structure (cfr. eq.  14 of Bel \& Marinoni 2012).
What is interesting about this statistic is that the
galaxy clustering ratio and the mass clustering ratio in real space,
which is defined as

\begin{equation}
\eta_{R}(r, {\bf p})\equiv \frac{\xi_R(r)}{\sigma_R^2} =
\frac{\int_0^{+\infty}\Delta_L^2(k, {\bf p}) \hat{W}^2(kR)
  \frac{sin(kr)}{kr} d\ln k}{\int_0^{+\infty}\Delta_L^2(k, {\bf
    p})\hat{W}^2(kR)d\ln k} ,
\label{core}
\end{equation}
coincide, or $\eta_{g,R}(r,{\bf p})=\eta_R(r,{\bf p})$.  This equality follows from the hypothesis that galaxy 
and mass density fields on a scale $R$ are related via a linear local deterministic biasing
scheme \citep{FG}.
Note that, $\hat{W}$ is the Fourier transform of the smoothing
window and $\Delta_L^2$ is the dimensionless linear power spectrum
of matter density fluctuations in Eq.~(\ref{core}).
Two fundamental observations motivated the definition of the galaxy
clustering ratio. Because it is a ratio between clustering at different scales, the clustering ratio is effectively 
insensitive to linear redshift distortions, which 
alter the amplitude of clustering in a way that is independent of
phase or frequency \citep{Kaiser}. We therefore have that

\begin{equation}
\tilde\eta_{g,R}(r, {\bf p})\equiv  \frac{\tilde \xi_{g,R}(r)}{\tilde \sigma_{g,R}^2}=\eta_R(r,{\bf p}),
\label{corezl}
\end{equation}
where quantities that are evaluated in redshift space are labelled
with a tilde. Because the amplitude of the galaxy clustering is not affected by galaxy bias,
we use this as the second salient feature of our definition.

The galaxy clustering ratio depends on the cosmological model through
both the linear power spectrum in Eq.~\ref{core}, which alters the
right-hand-side (RHS) of Eq.~(\ref{corezl}) and the conversion applied
to convert redshifts into distances, which alters the left-hand-side
(LHS). 
The equivalence expressed by Eq.~(\ref{corezl}) holds true if
and only if the LHS and RHS are both estimated in the `true'
cosmological model. This is the basis of the `Alcock-Paczynski' 
approach to constraining the cosmological geometry \citep{ap,bph,mb}.

There are a few  difficulties for the
 application of the clustering ratio test  to the VIPERS data. 
 The high density  and small volume  of the  VIPERS survey mean that
 we can only accurately
 calculate the $\eta$ statistic on smaller
 scales (typically $R=5\mpcoh$) when compared to those for which this
 cosmological indicator was originally conceived. 
 It is therefore necessary to
 extend the theoretical formalism of the $\eta$ test to
 account for the effects of structure formation in the
 non-linear regimes.  Adopting a non-linear power spectrum to predict
 the amplitude of the clustering ratio on these scales is necessary,
 but not sufficient.  We must also properly incorporate a model for the non-linear redshift space 
 distortion effect  induced by the  virial motions of  galaxies into the
 theoretical framework, or the
 so-called Finger-of-God effect. The additional modelling required  
 to extend the linear formalism of BM13 into the quasi-linear regime 
 is presented  and tested using  simulations of the large scale structure
 in  \S 4. Another concern is that  results obtained on small
 scales might be more sensitive to a failure of a fundamental
 hypothesis on which the $\eta$ formalism is built, or the locality
of the biasing relation. Particular care is thus
devoted to demonstrate via the analysis of  numerical simulations   
that even the amplitude of the 
galaxy clustering ratio is not affected by using VIPERS galaxies as
tracers on such small scales as $R=5\mpcoh$. That is, it  can be safely  predicted  without requiring  the
specification of a biasing scheme.  

\subsection{Estimating $\tilde \eta_{g,R}$  from VIPERS data}

The estimation of the clustering ratio
as a counts-in-cells statistic was presented in BM13. Here we
review the measurement and describe the application to VIPERS data.
$\delta_N=N/\overline{N}-1$, where $N$ is the number of galaxies in spheres
of radius $R$ and $\overline{N}$ its average value (We compute its value at
the radial position ${\bf r}$, corresponding to some look-back time t
by averaging the cell counts within survey slices $r \pm 10R$).  These
spheres tessellate the whole VIPERS survey in a regular way; their
centres (called hereafter seeds) being located on a lattice of
rectangular symmetry and spacing $R$ (See Fig.~\ref{w1w4}).  The
1-point, second order moment $\smash{\sigma_{g,R}^2}$ is measured as the
variance (corrected for shot noise effects) of $\delta_N$.  The
spheres partially include gaps in the survey due to the spacing of the
VIMOS quadrants and to failed observations.  Additionally,
the sampling rate of the survey varies with the quadrant, as described in
Paper I.  The impact of these effects is investigated in \S 5.2.

To estimate the correlation function of the counts, we add a motif of
isotropically distributed spheres around each {\it seed} (See right hand side of
Fig.~\ref{w1w4}), and as {\it proper seeds} we retain only those for
which the spheres of the motifs lie completely within the survey
boundaries. The centre $j$ of each new sphere is separated from the
proper seed $i$ by the length $r=nR$ (where $n$ is a generic real
parameter usually taken without loss of generality to be an integer),
and the pattern is designed in such a way as to maximize the number of
quasi-non-overlapping spheres at the given distance $r$ (the maximum
allowed overlapping between contiguous spheres is $2\%$ in
volume). Incidentally, if the galaxy field is correlated on a scale
$r=3R=15\mpcoh$, the number of spheres in a motif, or spheres
isotropically placed around each proper seed is 26. $\xi_{g,R}(r)$ is
then estimated as $\langle\delta_{N_i} \delta_{N_j}\rangle$, which is by averaging the
counts over any cell $i$ and $j$. In this last statistical quantity, note that we do not need to correct
for shot noise since random sampling
errors are uncorrelated. In \S 5.2, we discuss residual systematic
effects arising from the survey geometry and sampling rate.

Considering cells of $R=5\mpcoh$ and after removing those cells that
fall in bad or masked areas (see \S 5.2), the effective number of
spheres used to estimate 1-point statistics is $68,667$ in $W1$ and
58,684 in $W4$. Where 2-point statistics are concerned, the number of
proper seeds is $37,814$ in $W1$ and 39,459 in $W4$.  Note that we can
place more cells in $W1$ than in $W4$, since this last region is
characterised by a smaller field of view. On the contrary, the number
of proper seeds is larger in $W4$. This is explained by the angular
shape of the field. The shallower extension in declination of $W1$
limits the number of motifs that we can fit inside the survey area.
Since the error budget in the measurement of $\eta_{g,R}$ is
essentially dominated by the errors of the 2-point statistic, the
clustering ratio is thus estimated in $W4$ with slightly higher
precision. In this way, the small aperture of $W1$ also limits the
effective redshift range we can probe. To be able to place an entire motif (see right hand side of top panel of figure \ref{w1w4} ) in $W1$, we need to restrict the analysis to redshifts above $0.65$.

\section{The clustering ratio in the non-linear regime}

Because of the survey geometry, the
galaxy clustering ratio can be precisely estimated from the VIPERS
data only 
on relatively small
scales, $R<8\mpcoh$. 
It is thus necessary to generalize Eq.~(\ref{corezl}) to 
non-linear regimes and to test the validity of the hypothesis
underlying such modelling.
This entails additional assumptions with respect to the simple
requirement of a deterministic biasing relation on a
given smoothing scale $R$. 
Two distinct distortions must be modelled:
linear theory over predicts the real space amplitude of the galaxy clustering ratio
$\eta_{g,R}$ in regimes of strong gravity, and the redshift-space amplitude 
of the galaxy clustering ratio $\tilde{\eta}_{g,R}$ is biased high with respect to $\eta_{g,R}$ 
because of the Finger-of-God (FoG) effect \citep{jackson72,T&F}. Concerns about these points can be 
addressed by using numerical simulations. 
We create these from the
MultiDark simulation \citep{Prada11}, a
large flat $\Lambda$CDM simulation containing $2048^3$ particles, with
mass of $8.721 \times 10^{8} \msunoh$ within a cube of side
$1\gpcoh$. The simulation starts at an initial redshift $z_i=65$
with the following parameters:
$(\Omega_m=0.27,\; \Omega_\Lambda=0.73,\; H_0=70\kmsmpc,\;
\Omega_b=0.0469, \; n_s=0.95,\; \sigma_8=0.82)$.  We match VIPERS
against the time snapshot corresponding to $z=1$ (although the
sense of our conclusions is unchanged when other time outputs are
considered in the analysis).  This box (hereafter labelled $B_h$)
contains nearly 14 millions haloes with mass
$M>10^{11.5}h^{-1}$M$_{\odot}$ and is used to check real-space
properties of the clustering ratio with high statistical resolution.

We also consider a suite of 31 nearly independent light-cones, each
extends over the redshift range $0.65<z<1.2$ and covers a
cosmological volume similar to that surveyed by the VIPERS PDR-1 data
in the $W4$ field.  These light-cones (hereafter indicated as $L_h$)
incorporate redshift distortion effects, contain a total of $\sim
10^5$ haloes each, and have a constant comoving radial density of
objects, which is $\sim 5$ times higher than the mean effective density
of galaxies in the VIPERS sample.  These samples allow us to analyse
the redshift space properties of the
$\tilde \eta_{g,R}$ observable with great statistical resolution.

Note that the simulated haloes do not cover the whole range of masses
in which the VIPERS galaxies are expected to reside, resulting in a
different bias with respect to the real data.   This is not expected to
affect the realism of the tests performed, given that the $\eta$
statistic is insensitive to linear galaxy bias. This aspect is 
discussed in depth in \S~\ref{sec:bias}.

\begin{figure}
\includegraphics[width=90mm,angle=0]{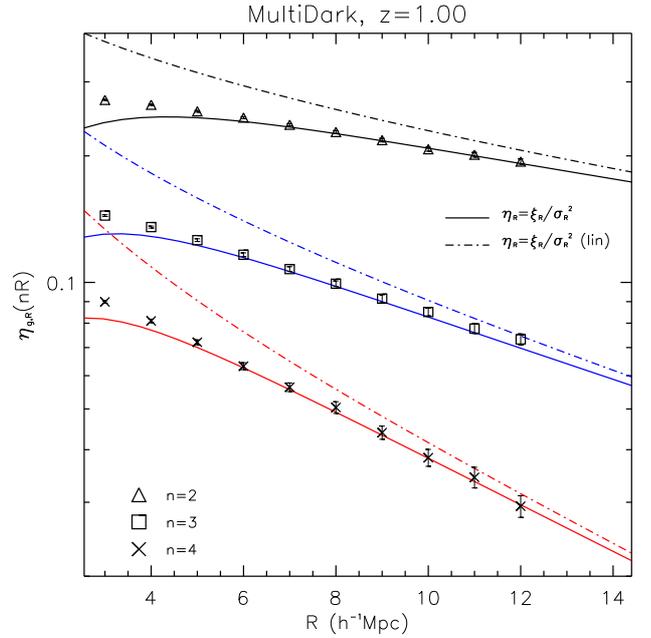}
\caption{The impact of non-linear clustering on the
  measurement and theoretical modelling of the galaxy clustering ratio $\eta$.  
The latter is measured in real space for the halo catalogues of the
$B_h$ simulation and compared to both the linear and non-linear
predictions, as described respectively by a simple \citet{EH} linear
power spectrum (dot-dashed lines) and after correction through the
non-linear {\it HALOFIT} prescription 
of \citet{Smith03} (solid lines). The clustering
  ratio is shown as a function of the smoothing scale $R$ adopted to
  filter the discrete distribution of haloes and for various
  correlation lengths ($r=nR$) with $n=\{2,3,4\}$.
Errors bars are obtained using $64$ block-jackknife resampling of the
data which excludes a cubic volume of linear size $250\mpcoh$ each time.
}
\label{figetanlr}
\end{figure}

\subsection{Non-linear effects in real space}

We first explore whether the assumption of a simple phenomenological
prescription for the amplitude and scaling of the matter power
spectrum provided by \cite{Smith03} is
accurate enough for an effective implementation of the $\eta$-test
with VIPERS data. We estimate
the galaxy clustering ratio in real space ($\eta_{g,R}(3R)$) using the numerical simulations and 
compare them against the model given by the RHS of Eq.~(\ref{core}),
which is
calculated using either the linear power spectrum, or the non-linear
model of \citet{Smith03}.
Figure~\ref{figetanlr} shows the failure of the predictions 
based on
a simple linear
power spectrum model
for the $B_h$ halo box at $z=1$.
The amplitude of the clustering ratio is systematically over predicted
on all $R-$scales and for all correlation lengths investigated.
In contrast, the simple non-linear model of \citet{Smith03}
describes the data with greater accuracy over a wider interval of
scales.  On a scale as small as $R=5\mpcoh$, the typical scale adopted
to extract the maximum signal from VIPERS data, or the precision with
which the amplitude of $\eta_{g,R}(3R)$ is predicted is of order
$3.6\%, \; 2.5 \%$ and $2.8\%$ for the correlation scales $n=2, 3$ and
$4$, respectively.  On scales $R=8\mpcoh$ and for the same correlation
lengths, the relative discrepancy between theory and data is of order
$1\%, \; 1.5\%$, and $2.5\%$, respectively.

These figures suggest that the remaining systematic error on the
real-space model calculated by adopting the \citet{Smith03} model 
is essentially negligible compared to the statistical uncertainty
associated with current measurements of the clustering ratio
$\eta_{g,R}$ from VIPERS data (which is, $\delta \eta/\eta
\sim 10\%$ at best when the field is filtered on the scale $R=5\mpcoh$).
Considering that the clustering ratio recovered on the correlation
scale $n=2$ and $n=4$ is affected respectively by too large
theoretical systematics (non-linearities in the power spectrum), by
too large observational uncertainties (due to the shape of VIPERS fields), and in agreement with the analysis of
BM13, we focus the rest of our analysis to the specific case, where the smoothed field is correlated on a scale $r=3R$.
Furthermore, we only consider the non-linear matter
power spectrum model of \cite{Smith03} from now on.

We remark that this analysis confirms the accuracy of the approximate
relation (\ref{corezl}) (upon implementation of a non-linear power
spectrum model) even in regimes where $\sigma_{g,R}$ is large.  This
is essentially due to the fact that, even on the small $R-$scales probed by VIPERS
the second order coefficient of the bias relation 
is small with respect to the linear term. 

\subsection{Non-linear effects in redshift space}

An interesting feature of Eq.~(\ref{core}) is that it is effectively insensitive to redshift distortions, and,
therefore, independent of their specific modelling in the linear
limit. 
This simplicity
is lost when we consider small scales, $R\sim5\mpcoh$, which are the typical
scales probed by VIPERS, where non-linear motions are
expected to contaminate the cosmological signal.

\begin{figure}
\includegraphics[width=90mm,angle=0]{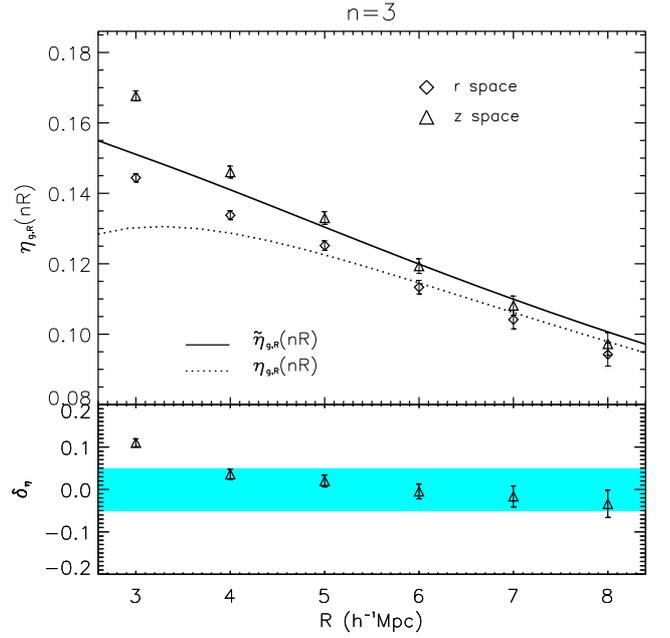}
\caption{\small  The effect of redshift-space distortions on the measurement
  and modelling of the clustering ratio.  The values estimated in real
  and redshift space from the $L_h$ halo catalogues (main panel)
 are compared with the theoretical predictions; these have been  
  named $\tilde \eta_{g,R}$ and $\eta_{g,R}$ respectively to indicate
  whether or not
  redshift-space distortions are included in the modelling.  Each
  point represents the average over $31$ mock catalogues (see text),
  while error bars give the corresponding standard deviation of the mean.
  The statistic is shown as a function of the filtering radius $R$ and
  correlates the field on scales $r=3R$.
To obtain $\tilde \eta_{g,R}$ from
  Eq.~(\ref{coreznl}) (solid line), we adopted a pairwise velocity dispersion
  $\sigma_{12}=200\kms$ \citep{bianchi,Marulli12}. The model accounts for both cases for non-linear
  evolution of the power spectrum by including the prescription of
  \citep{Smith03} into Eq.~(\ref{core}).  The lower panel shows
  instead the
  relative difference between the LHS and left- and right-hand sides
  of Eq.~(\ref{coreznl}) as a function of the same $R$. The 
  shaded corridor indicates a $\pm 5\%$ deviation.  }
\label{figetanlz}
\end{figure}

The existence of such effects is shown
in Fig.~\ref{figetanlz}, where we compare measurements of 
the galaxy clustering
ratio for haloes selected in real space and in redshift space from
the $L_h$ simulations ($\eta_{g,R}$ and $\tilde\eta_{g,R}$, respectively). The discrepancy
between measurements is significant with relative deviations
$\sim 14$\% for $R=4\mpcoh$ and $\sim 9$\% at
$R=5 \mpcoh$. 
In the following, we show that theory allows us to account for these
distortions in a neat and effective way. 

\begin{figure}
\includegraphics[width=90mm,angle=0]{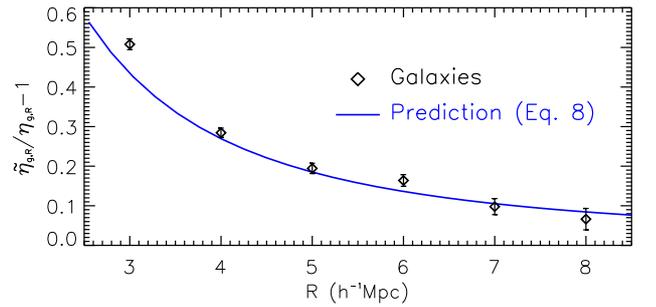}
\caption{\small Relative difference between the clustering ratio
    in real ($\eta_{g,R}$) and redshift ($\tilde\eta_{g,R}$)
    space. Black diamonds represent measurements from realistic galaxy mock catalogues  characterized by a small scale velocity dispersion $\sigma_{12}=360$ km s$^{-1}$. 
    The solid line shows  predictions obtained by inserting this specific value $\sigma_{12}$ into Eq.~(\ref{coreznl})}
\label{fig3b}
\end{figure}

A typical assumption is that the non-linear distortion induced on the $2$-point
correlation function is driven by random motions and can be described as a
convolution of the real-space correlation function with a Gaussian or
Exponential kernel $f(v_{12})$ that describes the distribution of pairwise
velocities along the line of sight (Davis \& Peebles 1983)

\begin{equation}
\tilde\xi(s_{\perp},s_{\parallel})=\int_{-\infty}^{+\infty}\dif v_{12}
f(v_{12})\xi \Big(s_{\perp},s_{\parallel}-\frac{1+z}{H(z)}v_{12} \Big) \,\,\,\, ,
\label{xirp}
\end{equation}
where $s_{\parallel}$ and $s_{\perp}$ are the separations along and
perpendicular to the observer's line of sight, $H(z)$ is the Hubble
parameter at redshift $z$, and
\[
f(v_{12})=\frac{1}{\sqrt{\pi}\sigma_{12}}e^{-\frac{v_{12}^2}{\sigma_{12}^2}}
\,\,\,\, .
\]
With this definition note that the 1D Gaussian pairwise velocity dispersion $\sigma_{12}$ is
$\sqrt{2}$ times  the dispersion in the pairwise velocity   $v_{12}$  and it induces
a dispersion in the radial comoving distance of amplitude
$\sigma_x=\sigma_v (1+z)/H(z)$,  where $\sigma_v=\sigma_{12}/2$ is  the dispersion of  galaxy peculiar velocities.
Interestingly, this kernel has also been shown to model quasi-linear redshift-space effects
\citep{per09}. This model can be straightforwardly re-mapped to Fourier space \citep{P&D94}, where the global
(linear coherent + non-linear random) redshift-space distortions
can be expressed as
$$
\tilde \Delta^2_{g,NL}(k)= \Delta^2_{g,NL}(k){\mathrm
  G}(k\sigma_x,\beta)\,\,\,\, .
$$
In this expression, $\tilde{\Delta}^2_{g,NL}(k)$ and ${\Delta}^2_{g,NL}(k)$ give the (non-linear)
power spectrum in redshift and real space, respectively, and

\begin{equation}
\begin{array}{rl}
G(y,\beta)\equiv &  \displaystyle\frac{\sqrt{\pi}}{8} \frac{{\rm erf}(y)}{y^5}\left[ 3\beta^2+4\beta y^2+4 y^4\right] \\
           & \displaystyle - \frac{e^{-y^2}}{4 y^4}\left[
             3\beta^2+2\beta(2+\beta)y^2\right] \,\,\,\, 
\end{array}
\label{pred}
\end{equation}
with $\beta=f/b$ as the usual redshift space distortion parameter
defined as the ratio between the linear growth rate and the linear
bias parameter.  Since we are essentially interested on scales, where
$y < 1$, we can expand Eq.~(\ref{pred}) as

\begin{equation}
G(y,\beta)\simeq K\left\{1-B_2\frac{y^2}{3}+B_4\frac{y^4}{10} - B_6\frac{y^6}{42}.....\right\},
\label{aprox}
\end{equation}
where, 
$$
\begin{array}{rl}
K   & = 1 + \frac{2}{3}\beta+\frac{1}{5}\beta^2 \\
B_2 & = 1 + \frac{2}{5}K^{-1}(4/3\beta+4/7\beta^2)\\
B_4 & = 1 + \frac{4}{3}K^{-1}(4/7\beta+4/15\beta^2)\\
B_6 & = 1 + K^{-1}(8/9\beta+24/55\beta^2).
\end{array}
$$ 
$K$ is the familiar Kaiser correction factor,
while the even coefficients $B_{2n}\sim 1$ in the limit $\beta < 1$.  Under these conditions,
\begin{equation}
G(y,\beta)\simeq KG(y,0).
\label{aproxbis}
\end{equation}

With this approximation and $G$ computed from Eq.~(\ref{pred}), we can predict
with sufficient accuracy  the non-linear amplitude of $\eta$  in redshift
space, 

\begin{eqnarray}
\tilde \eta_{g,R}(r, {\bf p}) &=& \tilde \eta_R(r,{\bf p})  \label{coreznl} \\
     &=&  \frac{\int_0^{+\infty}\Delta^2_{NL}(k, {\bf p})
       G(k\sigma_x,0)\hat{W}^2(kR) \frac{sin(kr)}{kr} d\ln
       k}{\int_0^{+\infty}\Delta^2_{NL}(k, {\bf
         p})G(k\sigma_x,0)\hat{W}^2(kR)d\ln k}. \nonumber  
\end{eqnarray}

Interestingly, both the linear bias parameter  and  the scale-independent Kaiser correction have dropped out of
this expression with the above ``derivation'' isolating the leading order contribution  in $G(y,\beta)$
and at the same time, decoupling the Kaiser model from the random dispersion.
For a realistic value of $\beta$ ($\simeq
0.5$), assuming $\sigma_v=175\kms$ and considering the scale
$R=5\mpcoh$, the approximation (\ref{coreznl}) is
$1.6$\% at $z=1$.

Figure~\ref{figetanlz} shows how well Eq.~(\ref{coreznl}) describes the
actual behaviour of haloes in the $L_h$ catalogues. For $R=4\mpcoh$ the relative discrepancy
between theory and data is reduced from $14\%$ to $4\%$, while for 
$R=5\mpcoh$ is as small as $2\%$.  As discussed in \S 4.1, this level
of systematic error (which is less than
the statistical errors to be expected from VIPERS) is expected from
the limitations of the adopted phenomenological description of the
non-linear power spectrum in Eq.~(\ref{coreznl})
\citep{Smith03} or from a possible non-local nature of galaxy bias at small $R$.
The accuracy achieved with this simple model over the wide range of
filtering scales $R$ is remarkable, especially because it
is not the result of fine-tuning the pairwise dispersion $\sigma_{12}$.
In Fig.~\ref{figetanlz}, the value was fixed to the value
$\sigma_{12}=200\kms$ as measured from
a dark matter simulation with similar mass resolution
\citep{bianchi,Marulli12}.  
Such a low value for $\sigma_{12}$ is expected since haloes do not sample their own inner velocity dispersion. In \S 5.3, we discuss
the robustness of Eq.~(\ref{coreznl}) in more detail for a wider range of velocity
dispersions. Notwithstanding, we show in Fig.~\ref{fig3b} how well the simple
 model  given in Eq. (8) accounts for small scale peculiar velocities in more realistic
mock catalogues. To this purpose, we consider a galaxy sample obtained by  
populating haloes according to the halo occupation distribution (HOD)
method. In \S 5.1 of \citet{delaTorre}, they describe step by step
how they obtained these mock catalogues from the MultiDark
simulation \citep[see][]{Prada11}. 

These results motivate our choice of $R=5\mpcoh$ as 
a smoothing scale for the analysis of the VIPERS data, since it maximises both the
accuracy (minimizing the distance between data and theory, see Fig.~\ref{figetanlz}) and the
statistical precision of the estimate of $\eta$. 

\begin{figure*}
\centerline{
\includegraphics[width=80mm,angle=0]{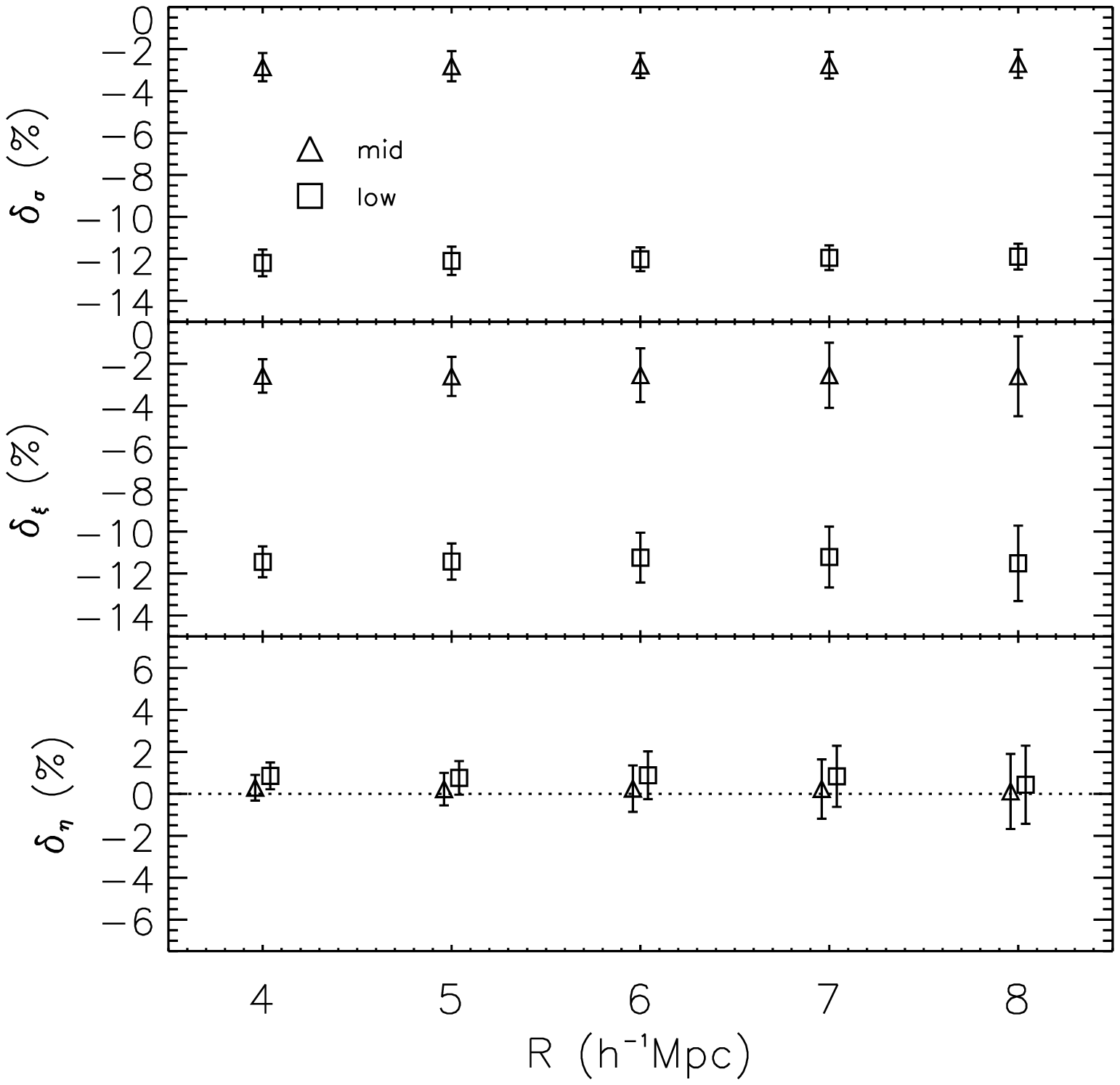}
\includegraphics[width=80mm,angle=0]{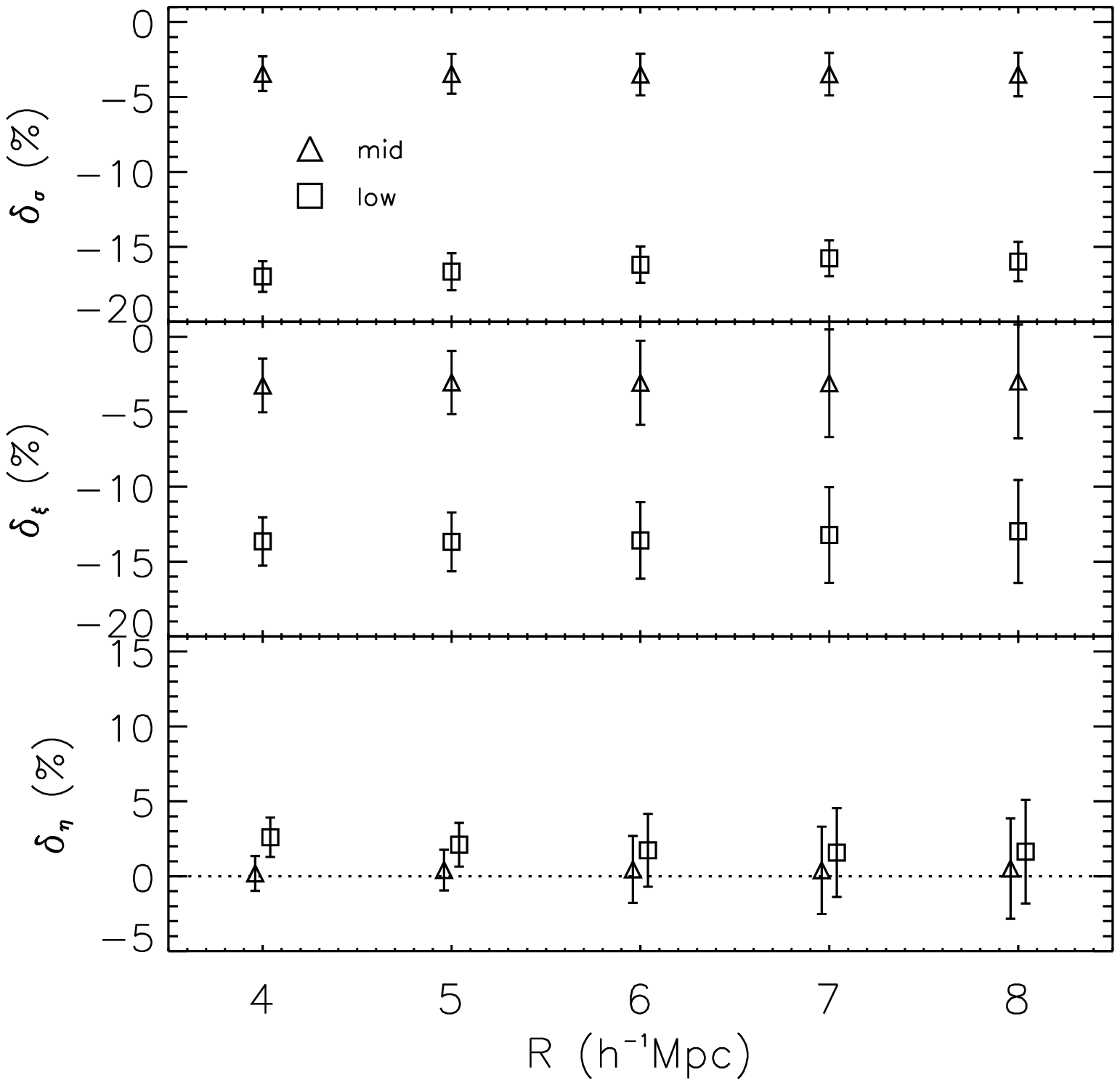}}
\caption{\small Testing the sensitivity of $\eta$ to the bias of the
  adopted tracers.  Clustering statistics are estimated from two
  sub sets of the halo simulated catalogues, known as {\it mid} and
  {\it low} sub-samples tant include haloes with masses limited to $M <
  10^{13}h^{-1}$M$_\odot$ and $M < 10^{12}h^{-1}$M$_\odot$,
  respectively.  The left and right panels show the relative variation in the variance
  $\delta_\sigma\equiv \sigma_{g,R}^2/\sigma_{g,R,ref}^2-1$ 
  (\textit{upper}), correlation function $\delta_\xi\equiv
  \xi_{g,R}/\xi_{g,R,ref}-1$ (\textit{central}), and clustering
  ratio $\delta_\eta\equiv \eta_{g,R}/\eta_{g,R,ref}-1$ with respect
  to the measurements from the  
  full mock catalogues that contain all haloes with masses up to $M = 7 \times
  10^{14}h^{-1}$M$_\odot$ for these two
  catalogues,. 
  In the left panel, quantities are measured from the $B_h$ halo
  catalogues in real space, while the measurements on the right are
  performed in redshift space using the $L_h$ light-cones.
  Error bars are estimated using the dispersion among the mock catalogues. For the sake of clarity, a relative shift
  along the abscissa is applied to the data in the bottom plots.}
\label{figbias}
\end{figure*}

\subsection{Sensitivity to galaxy bias}
\label{sec:bias}

In the linear regime, Eq.~(\ref{coreznl}) is independent of galaxy
biasing, where the clustering ratio does not depend on the particular
tracer we use to estimate it.  This is clearly only an approximation
when non-linear scales are included.  Evidence of the inadequacy  of the hypothesis of locality, 
where a scale-dependence of galaxy  bias ($\Delta_g^2(k)=b^2(k) \times \Delta^2(k)$), emerges naturally on
small scales.  Similarly,  the approximated redshift-space relation (\ref{coreznl}) 
explicitly neglects high order $\beta$ contributions.  
We now establish 
the scales where the $\eta$ statistics remains
independent of the choice of the galaxy tracer.

The effects of scale-dependent bias  are mitigated by  the very definition of the $\eta$  statistic  as a ratio.
For example, let's consider the biasing  model  $b^2(k)=(1+Q k^2)/(1+A k)\Delta^2_{L}/\Delta^2_{NL}$
with parameters $A=1.7$ and $Q=9.6$  (Cole et al 2005).  The  relative variation $db^2/b^2$ is as high 
as $32\%$ in the interval $0.01<k(h$/Mpc)$<1$, but results in $\eta_R(r)$  change by less then $7\%$.
Indeed, the clustering ratio on  scales $(R,r)=(5,15)h^{-1}$Mpc
roughly measures the relative strength  of the power spectra at  $k_{j_0}\sim0.09h$/Mpc  and $k_{W}\sim0.3h/$Mpc, as shown by BM13 (see \S 3 of that paper),
and, as a consequence,  it is  only sensitive to the variation in the bias between these two scales.
Although useful in appreciating the favourable properties of the $\eta$ statistic, 
this  simple argument is clearly insufficient  to quantify  the impact of scale-dependent bias of  VIPERS data on $\eta$.
Analysing  galaxy simulations is a more effective way to estimate the amplitude of  the remaining  systematic
error.

We use the $B_h$ (in real space) and $L_h$ (in redshift
space) simulated catalogues
at $z=1$, which contain haloes with masses up to $M= 7\times 10^{14}h^{-1}{\mathrm
M}_\odot$.  From these, we select two 
sub-samples that, have a lower bias compared to
the full parent catalogues, by construction.  We call these \textit{mid} and
\textit{low} samples and include, respectively, only haloes
with masses $M < 10^{13}h^{-1}{\mathrm M}_\odot$
and $M < 10^{12}h^{-1}{\mathrm M}_\odot$. The variance and $2$-point
correlation function of the smoothed density contrast $\delta_R$ from these samples are plotted in the upper
and middle panels of Fig.~\ref{figbias}, as a function of $R$. These are expressed in terms
of their relative difference with respect to the full parent catalogues.  The different
bias of the two samples is evident  
on both local and non-local
scales, as probed respectively by the two statistics $\sigma_{g,R}$
and $\xi_{g,R}$. As predicted by theory, the lower panel confirms that, both
quantities are biased in similar ways, such that the clustering ratio
is essentially insensitive to this: 
systematic errors are
limited to less than $1\%$ ($3\%$) in
real (redshift) space with this choice of the mass threshold.
  
\begin{figure*}
\includegraphics[]{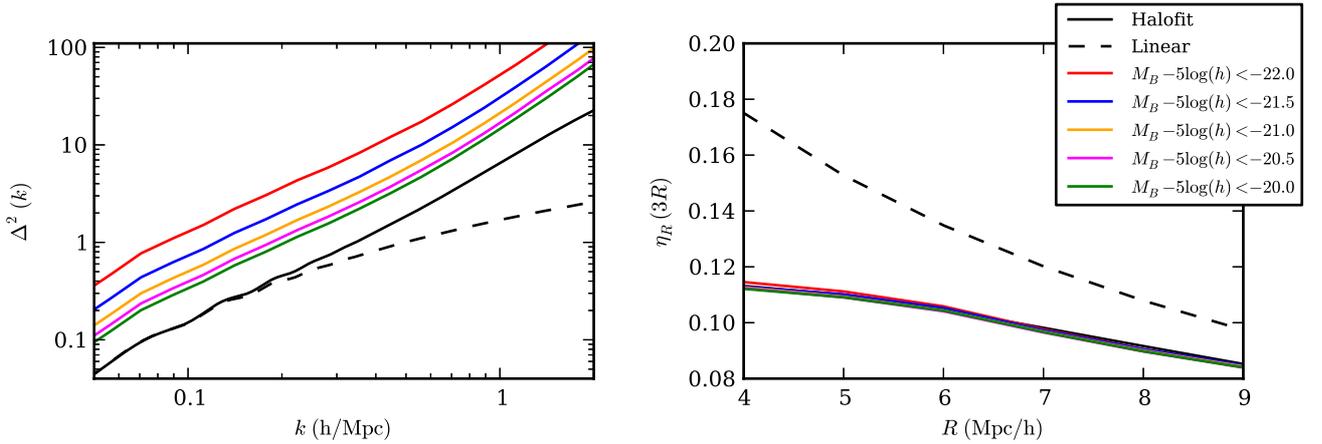}
\caption{\small {\it Left: } The best-fitting galaxy power spectra
  given the halo occupation distribution model fits for VIPERS
  galaxies in \citet{delaTorre}.  We examine five luminosity
  thresholded samples from $M_B<-22.0$ to $M_B<-20$ at z=0.8.  The
  {\it HALOFIT} matter power spectrum (black curve) and linear power
  spectrum (dashed curve) are overplotted.  The galaxy power spectra
  are consistent with a luminosity-dependent constant bias model with
  respect to {\it HALOFIT} up to $k=1 h/$Mpc.  {\it Right:} The values for
  the clustering ratio $\eta_R(3R)$ are computed for the VIPERS galaxy
  power spectra.  We find an agreement with the {\it HALOFIT} 
  prediction at the 2\% level at all scales down to $R=4 h^{-1}$Mpc.
}
\label{figviperspk}
\end{figure*}

A further indication of the limited impact of a scale-dependent bias
on the estimate of $\eta_R$ from the VIPERS data is provided by a
parallel analysis of the data
themselves. \citet{Marulli13} and \citet{delaTorre}
measure the dependence of the galaxy correlation function on
galaxy luminosity.  The presence
of any differential scale-dependent bias between samples with
different luminosities would make us
wary of non-local biasing effects entering in the clustering ratio as
well.  To investigate this, we consider the halo occupation
distribution model fits to the measured VIPERS two-point correlations
for five samples at $
\left <z\right >=0.8$ and luminosity between $M_B<-22$ and$-20$
\citep[Fig 12.]{delaTorre}. Figure~\ref{figviperspk} shows HOD model
power spectra 
$\Delta_g^2(k) = k^3P_g(k)/2\pi^2$ corresponding to the best-fit HOD 
parameters for these samples.  The plot shows that 
the shapes of the power spectra all follow the {\it HALOFIT} prescription, and only
begin to diverge at $k>1 h/Mpc$, or on scales to which $\eta_R$ has
little sensitivity. A luminosity-dependent
scale-independent bias model therefore provides an excellent match to the data. 
The clustering ratio is sensitive to the power spectrum over two
scales probed by the smoothed correlation function and the variance:
$\hat{W}^2(kR)j_0(nkR)$ and $\hat{W}^2(kR)$.  The range of wavenumbers
that affect the ratio is determined by the broader top hat kernel
$\hat{W}^2(kR)$ with effective width $k_{max}\sim \pi/R$.  From this
argument, we see that the clustering ratio is mainly determined by
the power spectrum at $k< 0.6 \hompc$ for $R=5$, thus by significantly
smaller $k$'s than indicated by the tests of Fig.~\ref{figviperspk}. 
This is explicitly shown by the right panel of the same figure, where 
$\eta_R(3R)$ is computed as a function of $R$ for the VIPERS galaxy
samples.  At $R=5\mpcoh$, the difference between the Halofit
prediction and the values estimated for the VIPERS samples differ by
no more than 2\%.  Given the $\sim 10\%$  random errors expected on the
estimates of $\eta$ from the current PDR-1 VIPERS sample (\S 6), 
this shows that any systematic effect due to scale dependence of galaxy
bias is expected to be negligible. 

\begin{figure*}
\centerline{
\includegraphics[width=90mm,angle=0]{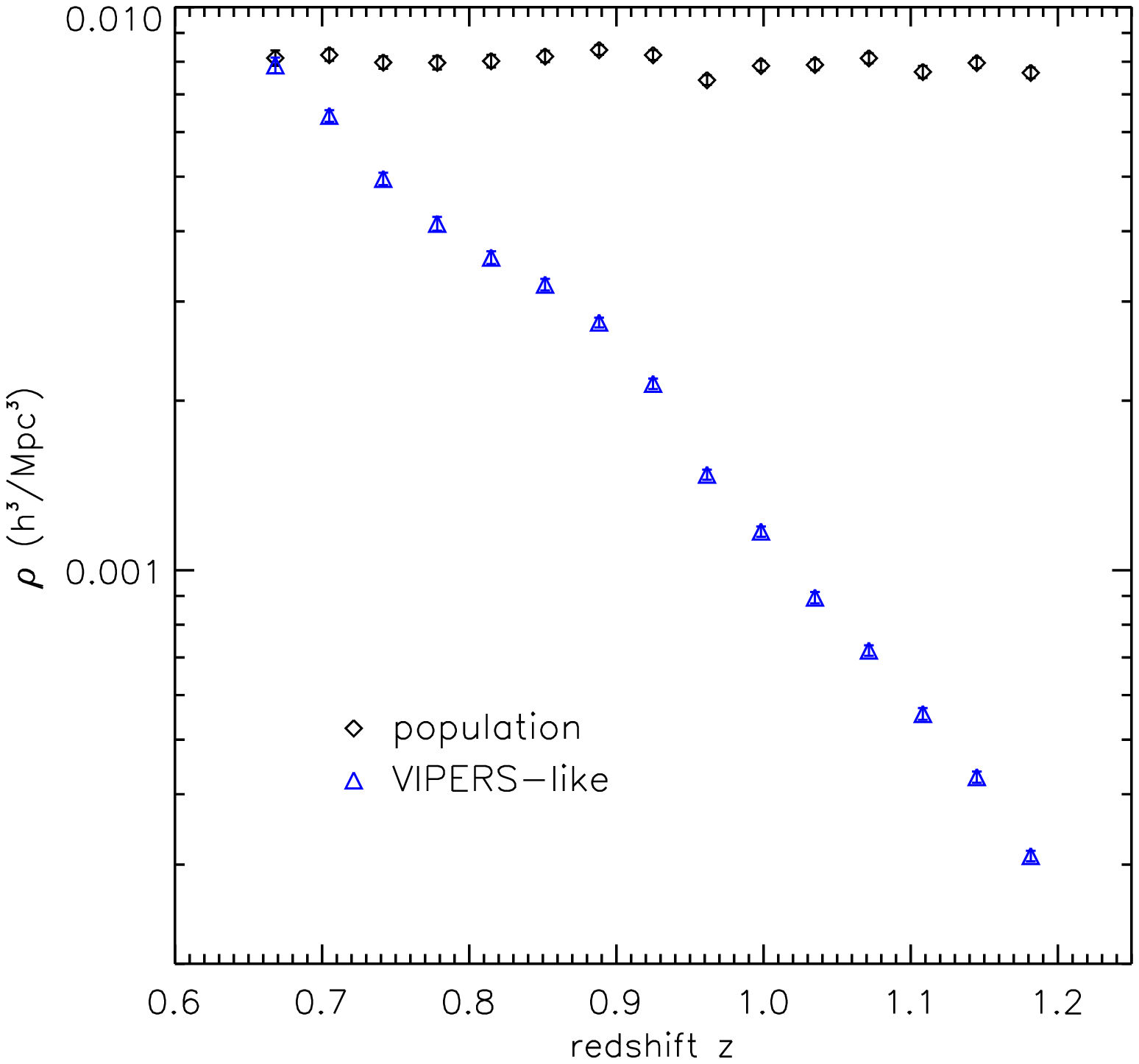}
\includegraphics[width=90mm,angle=0]{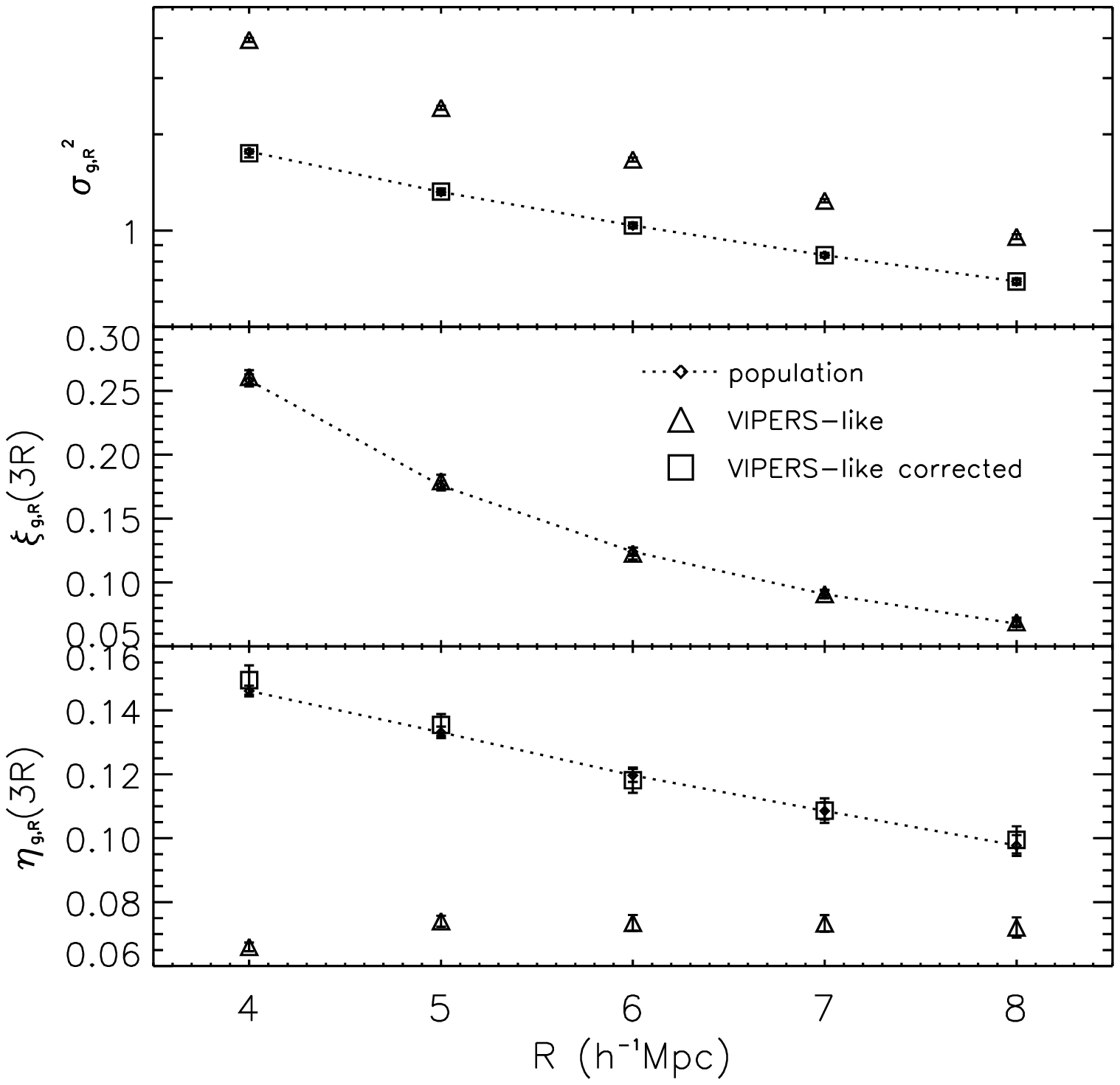}}
\caption{\small {\it Left:} The radial number density of galaxies in
  the simulated $W4$ field before (diamonds) and after (triangles)
  implementing the radial selection function of VIPERS PDR-1.
{\it Right:} Estimations of the variance $\sigma_{g, R}$ (upper),
  correlation function $\xi_{g,R}(3R)$ (centre), and clustering ratio
  $\eta_{g,R}(3R)$ (lower) from the true population of galaxies in the
  $W4$ field (diamonds) and from the sub-sample with the same VIPERS
  radial density profile (triangles) are shown as a function of the
  filtering scale $R$.  Squares show estimates obtained after
  correcting the relevant statistical quantities for shot noise,
  using the local Poisson model \citep{Lay}.}
\label{figden}
\end{figure*}

\section{The impact of observational effects on $\eta$}
\label{sec:obs-biases}
In this section, we discuss how observational effects have been
accounted for in our analysis, and we test the robustness and
limitations in producing an unbiased estimate of $\eta$ from the VIPERS data.

For this, we use 
simulated galaxy samples that implement the VIPERS selection effects.
We construct artificial galaxy light cones (named $L_g$ hereafter) by populating
the $L_h$ simulations with a specific Halo Occupation Distribution
(HOD) prescription, which are calibrated using VIPERS observations \citep{delaTorre}. The
next step in obtaining fully realistic VIPERS mocks is to add the
detailed survey selection function.  The procedure that we follow is
the same discussed in \cite{delaTorre}: we first apply the magnitude
cut $i_{AB} < 22.5$, then compute the observed redshift by incorporating
the peculiar velocity contribution and a random component that reproduces
the VIPERS redshift error distribution.  We then add the effect of the colour selection on the radial
distribution of the mocks. The latter is obtained by depleting the mocks
at $z < 0.6$, so as to reproduce the Colour Sampling Rate (CSR, see Paper I). The mock
catalogues that we obtain are then similar to the VIPERS parent
photometric sample. We next apply the 
slit-positioning algorithm \citep[SPOC,][]{Bottini05} with the same setting as for the data. This
allows us to reproduce the VIPERS footprint on the sky; the
small-scale angular incompleteness is due to spectra collisions and the variation in Target Sampling Rate across
the fields. Finally, we deplete each quadrant to reproduce the effect
of the Survey Success Rate (SSR).  In this way, we end up with a set of 31
and 26 realistic mock catalogues (named $L_{gV}$ hereafter), which
simulate the detailed survey completeness function and observational biases of VIPERS 
in the $W4$ and $W1$ fields, respectively.

It is important to remark that the parent mock galaxy catalogues used
here are not precisely the ``standard'' ones
created for VIPERS or those discussed in \citet{delaTorre}.  In the
latter, the mass resolution of the original $L_h$ halo catalogues was
improved by artificially adding low mass haloes following  the
prescriptions of \citet{delaTorre12}.  
This method uses the initial halo density field 
estimated on a regular grid, where the size is set by the mass resolution of 
the dark matter simulation. For the purpose of extending the halo mass 
range in the MultiDark simulation and creating VIPERS mock galaxy 
catalogues, an optimal grid size of $2.5h^{-1}$Mpc has been chosen. 
However, as shown in \citet{delaTorre12}, the grid size, or reconstruction 
scale $\lambda$, can have some impact on the accuracy with which galaxy two-point 
correlations are reproduced. In particular, values of $\lambda$ larger than the 
typical halo-halo separation, which are about $1h^{-1}$Mpc, can lead to an 
underestimation of the two-point correlation function of galaxies that 
populate the reconstructed haloes. In \citet{delaTorre12}, they showed that the two-point correlation function on scales around $1h^{-1}$Mpc is
underestimated by few percents for the faintest galaxies for 
the adopted reconstruction scale used to create the standard VIPERS mocks. We found that the 
variance of the smoothed galaxy field measured in the standard mocks is also 
affected by this effect. For this reason and without loss of generality, 
we thus decided to use the $L_g$ mock catalogues without the galaxies residing 
in the reconstructed haloes. Because these galaxies are 
hosted by haloes, which are systematically more massive than those hosting real 
VIPERS galaxies, does not affect the amplitude of the clustering ratio, which, as 
discussed in \S 4.3, is independent of the specific mass tracer.

\begin{figure*}
\centerline{\includegraphics[width=180mm,angle=0]{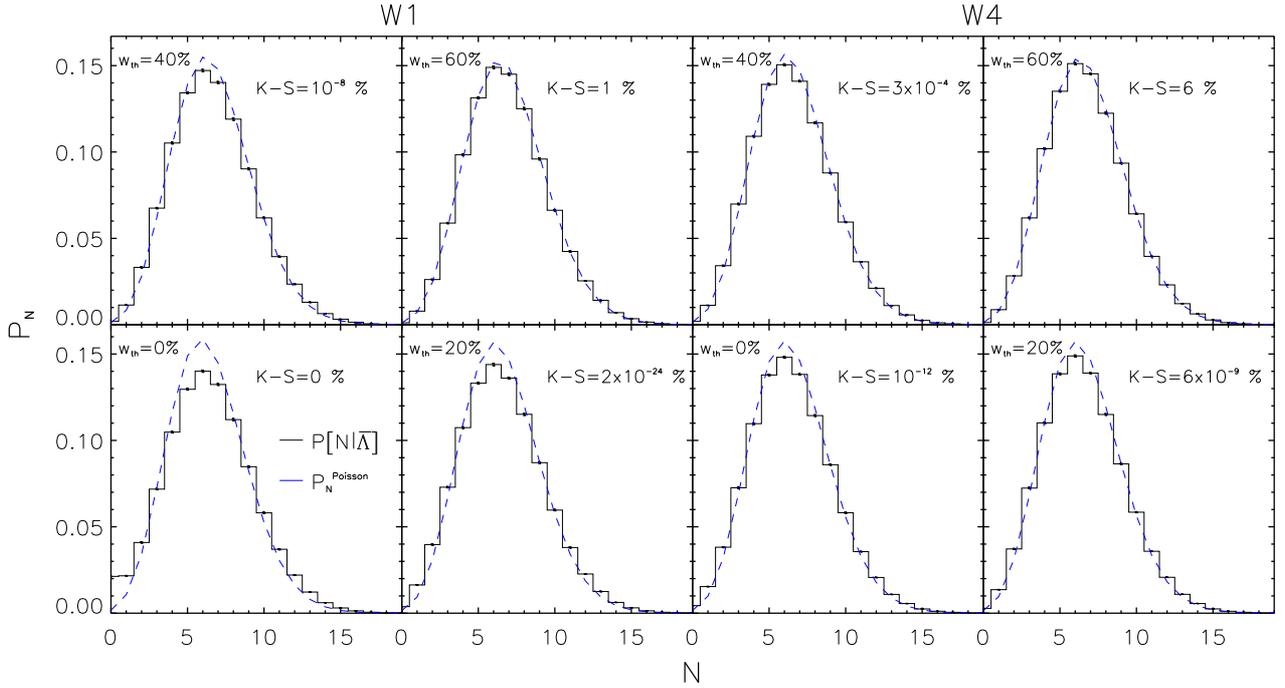}}
\caption{\small  The PDF of the number 
  count overdensities $\delta_N$ within spheres of radius
  $R=5h^{-1}$Mpc (histograms) are measured from mock surveys that reproduce
  the full selection functions of the VIPERS W1 and W4 fields (left
  and right groups of four panels). This is compared with a reference sample drawn
  from the Poissonian PDF shown by the dashed lines.  The four
  panels of each VIPERS field correspond to the PDF that one obtains when 
  rejecting an increasingly larger number of cells, which depends on
  whether the fraction of the cells volume affected by the survey mask
  is smaller than the threshold $w_{th}$ indicated in the insets.  The
  effect of ``corrupted'' cells is stronger in the W1 field, where it is necessary to reject all cells for
  which more than 40\% of the volume is affected by the survey mask to
  recover the correct PDF. Error bars are 
  obtained as the standard deviation over 50 distinct random
  catalogues. In the inset, we also quote the significance level of the
  Kolmogorov-Smirnov test on the agreement of the two distributions.}
\label{figpn}
\end{figure*}

\subsection{Effects related to the radial selection function: shot noise}

One effect of the selection function in a flux-limited sample is the
increase in the shot noise as a function of distance due to the corresponding decrease of
the mean density.  One could correct for this 
by increasing the size of the smoothing window $R$, but this would remove
the ability to compare fluctuations on the
same scale at different redshifts. Rather, we assume that the data
represents a local Poisson sampling of the underlying continuous
density field and correct for this statistically
\citep{Lay} and verify the limits of this assumption using our
mock samples. 

The right panel of Fig.~\ref{figden} shows the
effect of the shot noise correction on the two-point
correlation function, the variance, and the clustering ratio.  
While the two-point function is insensitive to shot noise by
construction, the variance does need to be corrected for the
increasing Poisson noise, which is given by the inverse of the
average counts in the spherical cells $\left[\bar{N}^{-1}(z)\right]$.
When subtracted from the observed value, the effect is completely
removed (square symbols).

\begin{figure}
\centerline{\includegraphics[width=90mm,angle=0]{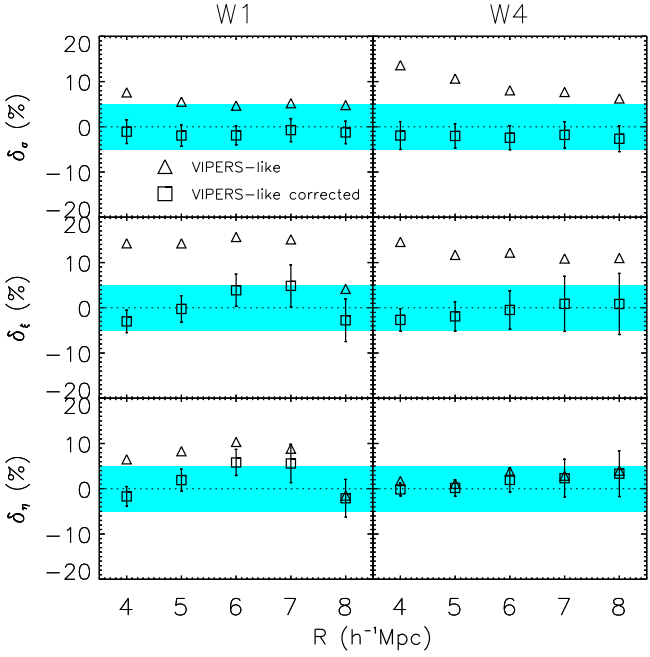}}
\caption{\small Testing the effectiveness of our correction for the VIPERS 
  angular selection function on the clustering ratio $\eta$. From top
  to bottom, we show (for a smoothing radius
$R=5h^{-1}$Mpc) the impact on the variance,  
$2$-point correlation function,
and clustering ratio $\eta_{g,R}(3R)$.  In each panel, we plot the mean and
scatter over the mock samples of the relative difference $\delta$ between
  the ``observed'' and ``true'' quantities (multiplied by 100):
  triangles are for no correction at all, while squares corresponds to
  measurements using the correction method discussed in the text. The
cyan shaded area correspond to a relative deviation smaller than $\pm
5\%$.  Left and right panels correspond to $W1$ and $W4$, respectively. 
}
\label{figmask}
\end{figure}

\subsection{Effects related to the angular selection function} 
\label{sec:angular}
The VIPERS redshifts are being collected by tiling the
selected sky areas with a uniform
mosaic of VIMOS fields. The area covered is not contiguous but
presents regular gaps due to the specific footprint of the instrument
field of view, as well as to intrinsic unobserved areas due to bright
stars or defects in the original photometric catalogue.  The VIMOS
field of view has four rectangular
regions of about $8 \times 7$ square arcminutes each, which are separated by an
unobserved cross \citep{Guzzo13, delaTorre}.
This creates a regular pattern of gaps in the angular
distribution of the measured galaxies, which is clearly 
visible in Fig.~\ref{w1w4}.  Additionally, the Target Sampling Rate
and the Survey Success Rate vary among the quadrants, and a few of the
latter were lost because of mechanical problems within VIMOS
(see Paper I for details). 
Finally, the slit-positioning algorithm, SPOC, also introduces some
small-scale angular selection effects   
with different constraints along the dispersion and spatial directions
of the spectra, as thoroughly discussed in \citet{delaTorre}.
Clearly, this combination of angular selection effects has to be taken
properly into account when estimating any clustering statistics.

In our specific case, this issue has been addressed as follows. We simulate a random (Poissonian)
distribution of galaxies (hereafter called ``parent catalogue'') and apply 
the global VIPERS angular selection function, which results from the
combination of the photometric and spectroscopic masks, the TSR and
the SSR. 

We test whether the distributions of counts within cells of radius $R$
match the expected Poisson distribution
in Fig.~\ref{figpn} using the ``parent'' $L_g$ (dashed
line) and ``observed'' $L_{gV}$ (histograms) mocks discussed earlier
on.  When the probability distribution function (PDF) is reconstructed using all possible
cells that can be accommodated within the rectangular footprint of the
$W1$ and $W4$ 
fields, the result is severely biased, which indicates that the
sampling of the underlying PDF is not random.  
It is clear that gaps and holes lead to 
a broadening of the PDF, which is artificially skewed towards 
low counts and an overestimate of the power in 1- and
2-point statistics.  We have demonstrated \citep{Bel13} that this
effect is mainly due to the missing full quadrants with a negligible
contribution of the smaller gaps produced by the VIMOS footprint, or
of the small-scale biases of the SPOC slit-positioning scheme.  We
also see that the effects are almost independent of the scale $R$ used
to compute the clustering statistics within the range explored here
($3-8 \hompc$).

\citet{Bel13} introduce a method
to correct for such angular selection effects and recover the correct shape of
the PDF and the corresponding moments to all orders. 
Here, we briefly summarize how we obtained a correct estimate
of the variance, which we need for building the clustering ratio. 
As shown by Fig.~\ref{figpn}, the size of the VIPERS PDR-1 sample is
such that we can simply  
reject all spherical cells for which more $40\%$ of the volume is
affected by the overall survey mask, which corresponds to regions not covered
by the survey.  With this threshold, we recover the random
sampling regime. 
Accordingly, we shall
reject all the ``seeds'' for which at 
least one sphere of the surrounding motif does not satisfy the
inclusion condition (see \S 3.2)  in the computation of the two-point function. Once this selection process is applied, 
the underlying statistical properties of galaxies are properly reconstructed without any additional de-biasing procedure. The net result on the estimated
statistics from the mocks is shown in Fig.~\ref{figmask}. The three
panels show that  
the variance, two-point correlation function, and clustering ratio for
the ``observed'' $L_{gV}$ mocks converge to the ``true'' value from
the parent $L_{g}$. Independent
of the correction, it is interesting to remark how, $\eta$ is fairly insensitive to these effects. This
happens because $\eta$ is defined as
the ratio of two quantities that are similarly affected.  
This is particularly impressive in the case of the $W4$ field, where $\eta$ is
virtually exact even without the correction.  
The price paid for this increased accuracy
is clearly a larger statistical error on $\eta$ due to the smaller effective
survey volume. 

\subsection{Impact of redshift  errors}

\begin{figure}
\includegraphics[width=90mm,angle=0]{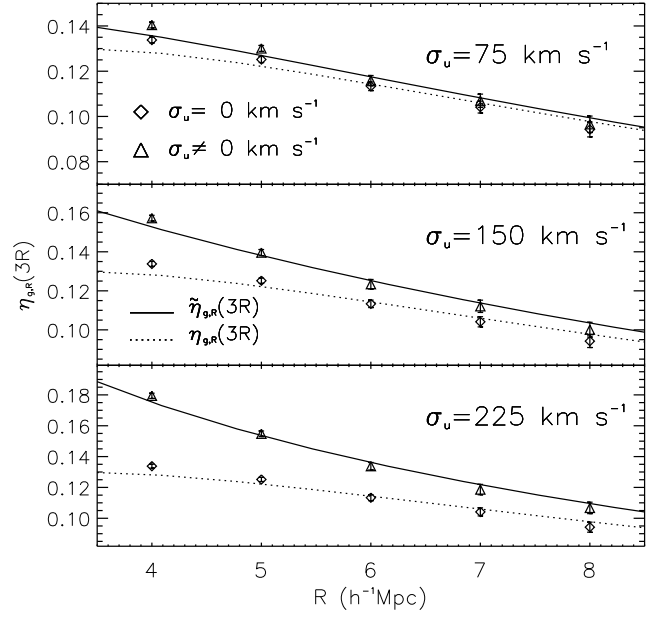}
\caption{The clustering ratio $\eta_{g,R}$ estimated from $L_h$ haloes
  in real space (diamonds) is shown as a function of the filtering
  radius $R$ and of the correlation length $r=3R$.  We also plot the
  clustering ratio $\tilde \eta_{g,R}$ estimated after perturbing the
  cosmological redshift of haloes with random errors
  (triangles). These random velocities are drawn from a Gaussian
  distribution with standard deviation $\sigma_{cz}=(1+z)\sigma_u$ km s$^{-1}$ with
  $\sigma_u$ as indicated. Also shown is the theoretically
  predicted value of $\tilde \eta_{g,R}$ obtained by inserting into
  Eq.~(\ref{coreznl}) the corresponding value of $\sigma_u$.
The points and error bars correspond to the average and standard
deviation of the mean of the measurements over $31$ $L_h$ catalogues.
Such error bars, in practice, would correspond to the typical
  uncertainty in a survey with $31$ times larger volume and 10 times
  higher galaxy number density than the VIPERS $W4$ sample.}
\label{figetazerr}
\end{figure}

\begin{figure}
\centerline{\includegraphics[width=90mm,angle=0]{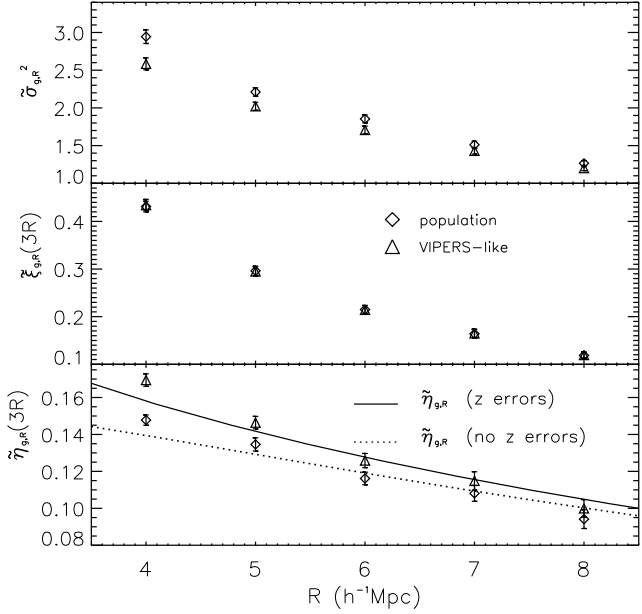}}
\caption{\small Impact and handling of the VIPERS redshift measurement errors on
  the usual statistical quantities that enter the definition of
  $\eta$, as plotted as a function of the filtering scale $R$ and 
 tested using the $L_g$ mock catalogues.  Triangles correspond to
  measurements performed 
  after adding a redshift error randomly
  drawn from a Gaussian distribution with standard deviation
  $\sigma_{cz}=141(1+z)$ to each galaxy position, which is representative of the errors
  affecting VIPERS redshifts \citep{Guzzo13}.  Diamonds give the
  reference values and are computed in redshift space without measurement
  errors. The dotted and solid lines give the clustering ratio
  predicted by Eq.~(\ref{coreznl}) when a dispersion
  $\sigma_{v}=100\kms$ or
  $\sqrt{\sigma_{v}^2+\sigma_{cz}^2(1+z)^{-2}}$ (i.e. adding in
  quadrature the expected redshift error) are used, respectively. 
  Expectation values
  and error bars are computed from the mocks in the usual way. 
}
\label{figetaztot}
\end{figure}

Using repeated observations of 1215 objects in the
the PDR-1 catalogue, the {\it rms} measurement error of VIPERS
redshifts has been estimated to be $\sigma_z=0.00047(1+z)$ or 
$\sigma_{cz}=141(1+z)$ km s$^{-1}$ (see Paper I).  At $z\sim 1$, this
translates into an error in the radial  
comoving distance of $\sim 3h^{-1}$Mpc, which is expected in principle
to have an impact on counts-in-cell statistics, for comparable cell
sizes $R$. 

The net effect of redshift errors
is to smooth the galaxy distribution in redshift-space along the
radial direction, suppressing the amplitude of 1-point statistics.
This is similar to the effect of small-scale random peculiar
velocities (although the latter on small scales are better described
by an exponential distribution rather than a Gaussian).  We therefore
model the effect of redshift errors in the expression of $\eta$ 
(Eq.~\ref{coreznl}) by using an effective
dispersion  $\sigma_{TOT}=\sqrt{\sigma_v^2+\sigma_{cz}^2(1+z)^{-2}}$ in the Gaussian damping term,
or by adding in quadrature the VIPERS {\it rms} redshift
error to the peculiar velocity dispersion of galaxies. 
We test the
goodness of this description using the $L_g$ mock surveys as described earlier,
to which we add a radial displacement drawn from a Gaussian distribution with
the appropriate dispersion.  The results of measuring the variance $\tilde\sigma_{g,R}$,
two-point correlation function $\tilde \eta_{g,R}$, and clustering
ratio $\tilde\eta_{g,R}$ of the smoothed
field on the perturbed and unperturbed mocks, as compared to the
corrected and uncorrected model, are shown in  Fig.~\ref{figetaztot}.
As expected, the figure shows that, as expected, the amplitude of the effect (an
underestimate of $\tilde\sigma_{g,R}$ and an overestimate of $\tilde
\eta_{g,R}$) increases for a decreasing smoothing scale $R$, when
the latter becomes comparable to the redshift errors. 
To test the correction, we have estimated $\eta_{g,R}$ in the $L_h$ light cones of haloes in real space
after having introduced random errors, which are characterised by the
dispersion parameter $\sigma_u$ ($\sigma_{cz}=\sigma_u(1+z)$), in the
cosmological redshift. The outcome of this analysis is presented in
Fig.~\ref{figetazerr} and shows that Eq.~(\ref{coreznl}) allows us to account for either small scale peculiar
velocities or redshift errors.

\subsection{Combined correction of systematic
  effects}

We finally want to compare how well the combination of the different pieces we have
developed and implemented into our description to account for
non-linear and observational biases is capable of recovering the
correct original value of $\eta$.  This test is performed by comparing
the ``idealized'' $L_h$ mock surveys, which contain a population of dark matter haloes with constant density (no
selection function, no mask) within the volume of the VIPERS
survey, and the set of $L_{gV}$ mocks,
which contain HOD simulated galaxies that are selected according to the
VIPERS selection function.  We want to test whether the clustering
ratio reconstructed from the VIPERS-like samples of galaxies after
the correction for all the survey selection functions traces the
clustering ratio reconstructed from the $L_{h}$ samples. In other
words, we want to determine if the clustering ratio is in agreement with
catalogues characterized by a different set of tracers (haloes)
masses (in the range $3 \times 10^{11}<M h^{-1} {\mathrm M}_{\odot}<7\times 10^{14}$)
 without any VIPERS observational selection function (except from
random redshift errors that we add to the haloes according to the
techniques explained in \S 5.3). 

\begin{figure}
\centerline{\includegraphics[width=90mm,angle=0]{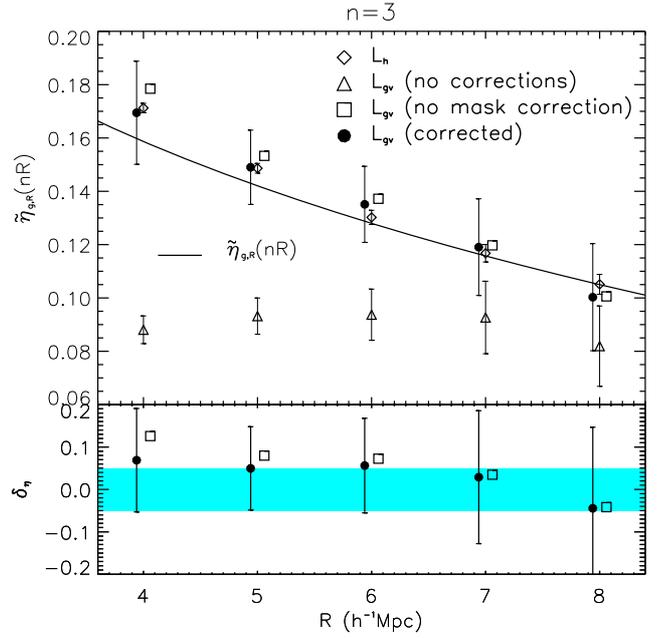}}
\caption{\small Overall impact and treatment of VIPERS selection
  effects on the estimate of the galaxy clustering ratio as a function
  of the smoothing scale $R$ and for a correlation scale $n=3$. The
  reference values of $\tilde \eta_{g,R}$ as measured
  in redshift space from the $L_h$ halo catalogues (diamonds) are
  compared with those estimated from the realistic VIPERS-like mock
  samples $L_{gV}$ which include the radial and angular selection
  functions.  The weight of the different corrections is determined by
  the different sets of points, as indicated in the legend.  As usual, the plotted points represent ensemble
  averages over the mocks. Error bars on the reference $L_h$ measurements
  (triangles) correspond to 
  the standard deviation of the mean, while errors from the simulated
  VIPERS mocks give a forecast of the actual errors expected from the
  VIPERS data analysed in this paper (including cosmic variance). The
  solid line gives the theoretically predicted value of the
  clustering ratio from Eq.~(\ref{coreznl}), including a pairwise
  velocity dispersion of $\sigma_{12}=200\kms$ (a plausible
  value, given the resolution of the simulation used
  \citep[see][]{bianchi, Marulli12}) and the {\it rms} error of the
  VIPERS redshifts \citep{Guzzo13}. Note that this theoretical
  prediction is given as reference and is not the best fit model,
  as given the data.} 
\label{fighod}
\end{figure}

The results of this analysis, as shown in
Fig.~\ref{fighod}, confirm the robustness of the clustering
ratio.  The same figure also shows the separate
impact of the different effects and how $\eta$ is particularly
insensitive to some of them, due to its definition, as already shown in the previous sections. 
Specifically, these are all those 
biases that affect 1- and 2-point clustering statistics in the same
sense.

\section{Cosmological constraints from the VIPERS PDR-1 data}

Using the methodology developed and tested in the previous sections, we are now in the
position to apply the clustering ratio test to the 
VIPERS survey and study how well we can constrain the values of (at
least some) cosmological parameters using the current catalogue.

%
%
\begin{figure}
\centerline{\includegraphics[width=90mm,angle=0]{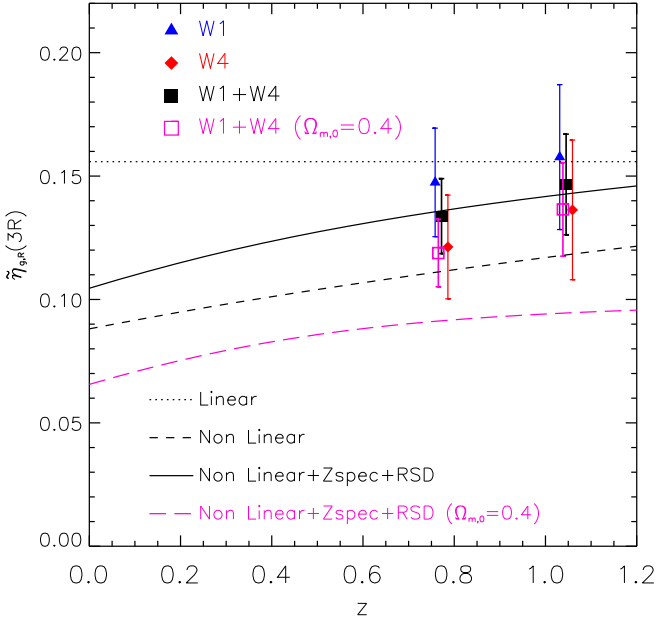}}
\caption{\small The galaxy clustering ratio $\tilde \eta_{g,R}$ on a
  scale of $R=5h^{-1}$Mpc and a correlation length $n=3$, as estimated from
  real VIPERS data (squares) in two distinct redshift
  bins. Measurements for the individual VIPERS fields $W1$ (diamonds)
  and $W4$ (triangle) are also shown.  The best fit theoretical
  model (cf. Eq.~\ref{coreznl}) accounting for all non-linear and
  observational effects is given by the solid line.  For
  comparison, we also display the corresponding linear theory
  prediction (dotted line) for the best fit cosmological
  parameters
  and the non-linear prediction when the corrections for peculiar
  motions and redshift 
  errors are not included (short-dashed line).  Error bars have been
  estimated from the standard deviation among the $31(/26)$ $L_{gV}$ mock VIPERS
  surveys of the $W1$/$W4$ fields, which include the
  contribution from cosmic variance.  We also plot the observed and
  predicted values of $\tilde
  \eta_{g,R}$ when a ``wrong'' cosmology with slightly higher matter
  density $\Omega_{m0}=0.4$ is assumed (open squares and long-dashed
  curve, respectively).  The divergence of observations and theory
  shows that this is not the best-fitting model, resulting in a low
  likelihood.  For clarity, all data points have been slightly shifted
  along the abscissa.} 
\label{figetagvsz}
\end{figure}

\subsection{Likelihood definition}

Let us first illustrate the procedure to evaluate $P({\bf
  p}|\eta_{g,R})$, which has the
likelihood of the unknown set of parameters ${\bf p}=(\Omega_m,
\Omega_{X}, w, H_0, \Omega_bh^2, n_s, \sigma_8, \sigma_{12})$, given
the observed value of $\eta_{g,R}$.
The probability distribution function of the clustering ratio is not
immediately obvious since this observable is defined via a ratio of
two non-independent random variables. By using both simulated
and real data, BM13 showed that the PDF of $\eta_{g,R}$ is very well
described by a Gaussian function, meaning that we can apply a standard
$\chi^2$ analysis.

In the analysis of the VIPERS sample we have considered two cases: 
\begin{itemize}
\item a single redshift bin 
  in the interval $0.65<z<1.2$ (i.e. $N=1$ in
  Eq.~\ref{coreznl});
  \item two uncorrelated bins, at $0.65<z<0.93$ and $0.93<z<1.2$, i.e. with $N=2$ in
Eq.~(\ref{coreznl}) (see
Fig.~\ref{figetagvsz}).
\end{itemize}
In the following, we shall take
their mean redshift as centres of these two bins rather than the median of the galaxy
distribution. This choice
is justified by the fact that this is approximately the redshift at
which the accuracy in the estimate of $\eta$, as determined by the
tradeoff between a decreasing number of objects and an increasing number of
cells, has its peak.  Note that this choice has no effect on the
cosmological results.

In our analysis, we assume a flat cosmology in which 
the accelerated expansion is caused by a cosmological
constant term in Einstein's field equations (i.e., we fix the dark
energy equation of state parameter to $w=-1$).  
As in BM13, we assume Gaussian
priors on  the baryonic matter density parameter $\Omega_b
h^2=0.0213\pm 0.0010$, the Hubble constant H$_0=73.8\pm 2.4\kms$Mpc$^{-1}$ and the primordial index 
$n_s=0.96\pm 0.014$, as provided respectively by BBN \citep{bbn}, HST \citep{hst} and WMAP7
\citep{lars} determinations. 
Additionally, we assume Gaussian priors on 
$\sigma_8$, which is
re-parameterized as $\Delta_R(k_{w_{map}})^2=(2.208 \pm 0.078)\times
10^{-9}$ with $k_{w_{map}}=0.027$ Mpc$^{-1}$ \citep{koma} and on
the effective pairwise velocity dispersion ($\sigma_{12}=2\sigma_{TOT}=514\pm24\kms$), as estimated from the VIPERS data themselves (private communication by de la Torre). 
Finally, we underline that we do not include the full WMAP likelihood, which would
introduce a strong prior on the matter density parameter $\Omega_mh^2$. 

The clustering ratio $\eta_{g,R}(r, {\bf p})$ is estimated from the data by re-mapping the observed
galaxy angular positions and redshifts into comoving
distances, which vary the cosmology on a grid in a $6D$ space
($\Omega_m$, $H_0$, $\Omega_b h^2$, $n_s$, $\sigma_8$, $\sigma_{12}$).
A consequence of this is that the posterior ${\mathcal L}$ does not vary smoothly
among the different models because the number of galaxies counted in any
given cell that varies from model to model. However, the computation of the observable
$\eta_{g,R}(3R, {\bf p})$ is rather fast, given the
counts-in-cells 
nature of the method.
Therefore, shot noise is the price we have
decided to pay in order to avoid fixing cosmological parameters at
fiducial values and to obtain an unbiased likelihood hyper-surface.

\begin{figure}
\centerline{\includegraphics[width=90mm,angle=0]{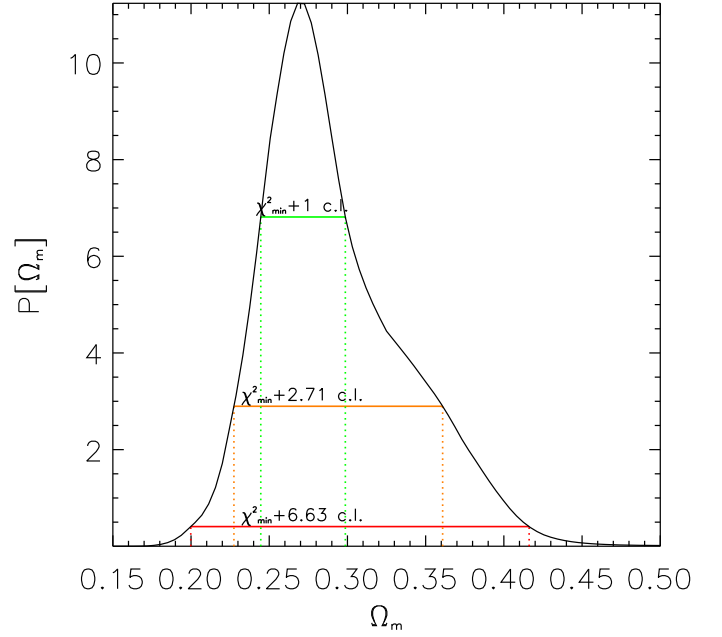}}
\caption{\small The normalized 1D posterior probability  of the density
  parameter at the current epoch, $\Omega_m$, estimated from the full
  VIPERS sample that is centred at $z=0.93$. The curve is obtained by
  marginalizing the posterior
  $P({\bf p}, \tilde \eta_{g,R}(3R)$ over the remaining ${\bf p}$
  parameters. The vertical dotted lines define the confidence
  intervals that correspond to $\chi^2_{min}+1$, $\chi^2_{min}+2.71$ and $\chi^2_{min}+6.63$.} 
\label{figlik}
\end{figure}

\subsection{Results}

In Fig.~\ref{figetagvsz}, we show the values of $\tilde
\eta_{gR}(nR)$ measured from the VIPERS data for
$(R,n)=(5h^{-1}$Mpc$,3)$, which splits the survey in different
fashions (points).  Note how the scatter between the estimates
obtained by analysing the two fields W1 and W4
separately is compatible with the error bars that are estimated from the
mock samples, suggesting that cosmic variance
largely dominates the error budget. The curves correspond to the model
(cf. Eq.~\ref{coreznl}), which is computed for the best fit values of the
parameters, as 
obtained for the whole survey. The solid line corresponds to the full
model, which accounts for all non-linear corrections, whereas
the dotted line gives the expected
redshift-independent behaviour if $\eta$ were measured (with the same
derived parameters) on
fully linear scales (for larger $R$ as, for example, in
BM13). The short-dashed line corresponds instead to only correcting
for non-linear clustering but not for redshift-space distortions,
while the magenta long-dashed line demonstrates the intrinsic sensitivity
to $\Omega_m$, which is further discussed in the following section (\S~\ref{sec:ap}).

The best fit parameters have been first estimated 
splitting the data into two redshift bins ($0.65 \le
z<0.93$ and $0.93<z \le 1.2$).  For these two volumes, the likelihood analysis yields
$\Omega_m=0.281^{+0.054}_{-0.043}$ and $\Omega_m=0.268_{-0.029}^{+0.034}$,
respectively. Noticeably, error bars are smaller in the high redshift
bin, despite the lower number density of galaxies. This is mainly the
result of the $30$\% larger total volume but also of the smaller fraction of
rejected cells in the process of mask correction.  When the 
full redshift range $0.65<z \le 1.2$ is analysed at once, we obtain
$\Omega_m= 0.270_{-0.025}^{+0.029}$, which is the value used to produce the
model curves in Fig.~\ref{figetagvsz}.  

Figure~\ref{figlik} shows the 1D
normalized posterior probability of $\Omega_m$ for this case, 
which is obtained by marginalizing $P({\bf p}, \eta_{g,R})$ over the remaining
${\bf p}$ elements.  It is also interesting to note that if each element of
the vector ${\bf p}=(\Omega_m, H_0, \Omega_bh^2, n_s, \sigma_8,
\sigma_{12})$ is perturbed by $+10\%$, the corresponding variation
induced in the amplitude of the clustering ratio is $(-8.5\%, -14.6\%,
2.7\%, 12\%, 3\%, 3\%)$.  This shows the specific sensitivity of
$\eta$ to the values of
$\Omega_m$, $h$, and $n_s$, as well as its weak dependence on 
$\Omega_b h^2$, $\sigma_8$ and $\sigma_{12}$.  The latter is
particularly relevant, as the last two parameters enter into play here
only because $\eta$ is pushed to non-linear scales. The indication is
therefore, once again, that the dependence on these specific
parameters is weak. By giving a $-40$\% change to the assumed central
value of the Gaussian prior on the peculiar velocity dispersion $\sigma_v$, we change
the estimated value of the matter density parameter by only
$-7$\%. This however remains as the biggest potential systematic in
our analysis.

\begin{figure}
\centerline{\includegraphics[width=90mm,angle=0]{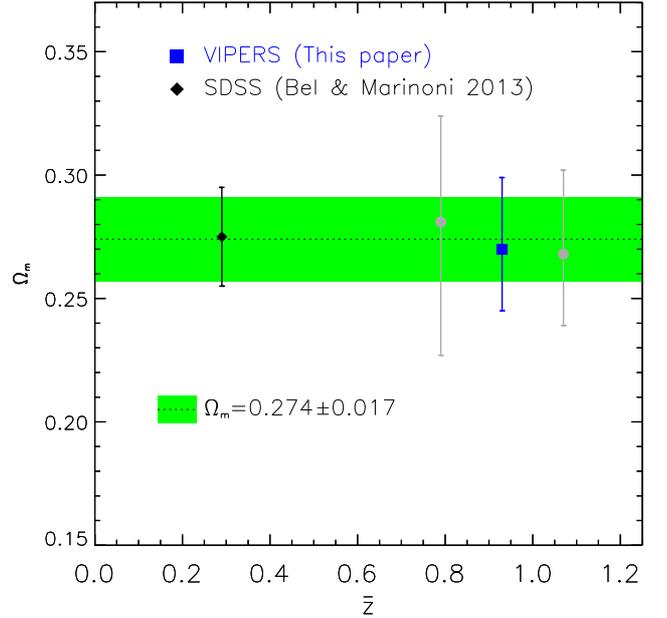}}
\caption{\small The value of the matter density parameter $\Omega_m$
  as estimated at different redshifts from the SDSS DR7 \citep{bm} and VIPERS
  samples; the full sample (blue square) and two non-overlapping redshift bins (grey filled circles) correspond to $0.65\le z<0.93$ and $0.93<z \le 1.2$. Both analyses assume a flat $\Lambda$CDM model and the same
  priors on common parameters. The dotted line shows the best fitting
  value for $\Omega_m$, as obtained combining the SDSS and VIPERS measurements.} 
\label{figlast}
\end{figure}

For comparison, the $\eta$-test to LRG galaxies of the
SDSS DR7 sample is applied and (but smoothing the field on larger, more linear scales using
$R=14h^{-1}$Mpc), BM13 obtain $\Omega_m=0.275\pm 0.020$.  This value
is compared to those obtained here in Fig.~\ref{figlast}.
The roughly comparable uncertainty of the VIPERS and SDSS
measurements, despite the smaller volume and number of objects of
VIPERS, 
is explained by the much larger number of cells available to
compute the statistic, when the field is smoothed on scales
$R=5h^{-1}$Mpc, as for VIPERS, rather than on 14$h^{-1}$Mpc, as used
in the SDSS analysis of BM13. The price to be paid in the former case is that of having to handle and model non-linear
scales.

%

The excellent consistency of these measurements over a wide redshift
interval ($0<z<1.2$) and from fully independent data sets is 
remarkable and strongly support the reliability of the measurements made
by \citet{delaTorre}. The combination of SDSS and VIPERS estimates of the matter density
parameter gives
$\Omega_{m,0}=0.274\pm 0.017$, which improves the precision
of the WMAP7 determination ($\Omega_m=0.267\pm0.029$, \citet{lars}) by $75\%$,
and by more than a factor of two when it is obtained by the joint
analysis of the WMAP3 and SDSS DR5 data ($\Omega_m=0.259\pm 0.039$,
\citet{eis}). Our precision is comparable to that obtained by \cite{per10}
($\Omega_m=0.278\pm0.018$) from the combination of BAO measurements
from SDSS DR7 and 2dFRGS and the full WMAP5
likelihood \citep{cmb5}.  We also note that the most recent analyses of the BOSS data
give $\Omega_m=0.298\pm 0.017$, when using BAO and WMAP7 
\citep{anderson12} and $\Omega_{m0}=0.282 \pm0.015$ from the
combination of the full correlation function with WMAP7 \citep{sanchez12}.  

\begin{figure*}
  \centerline{\includegraphics[width=90mm,angle=0]{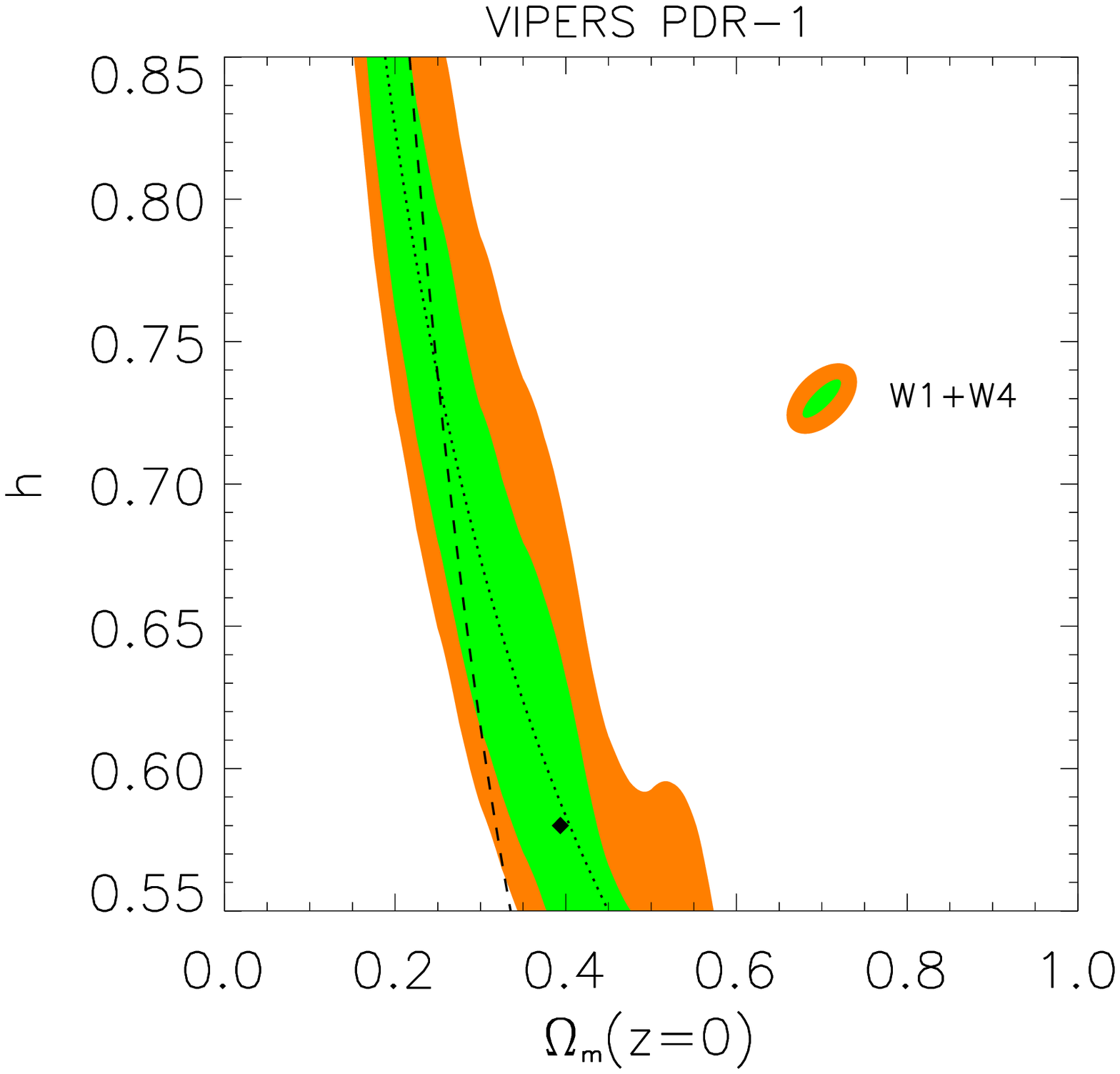}
    \includegraphics[width=90mm,angle=0]{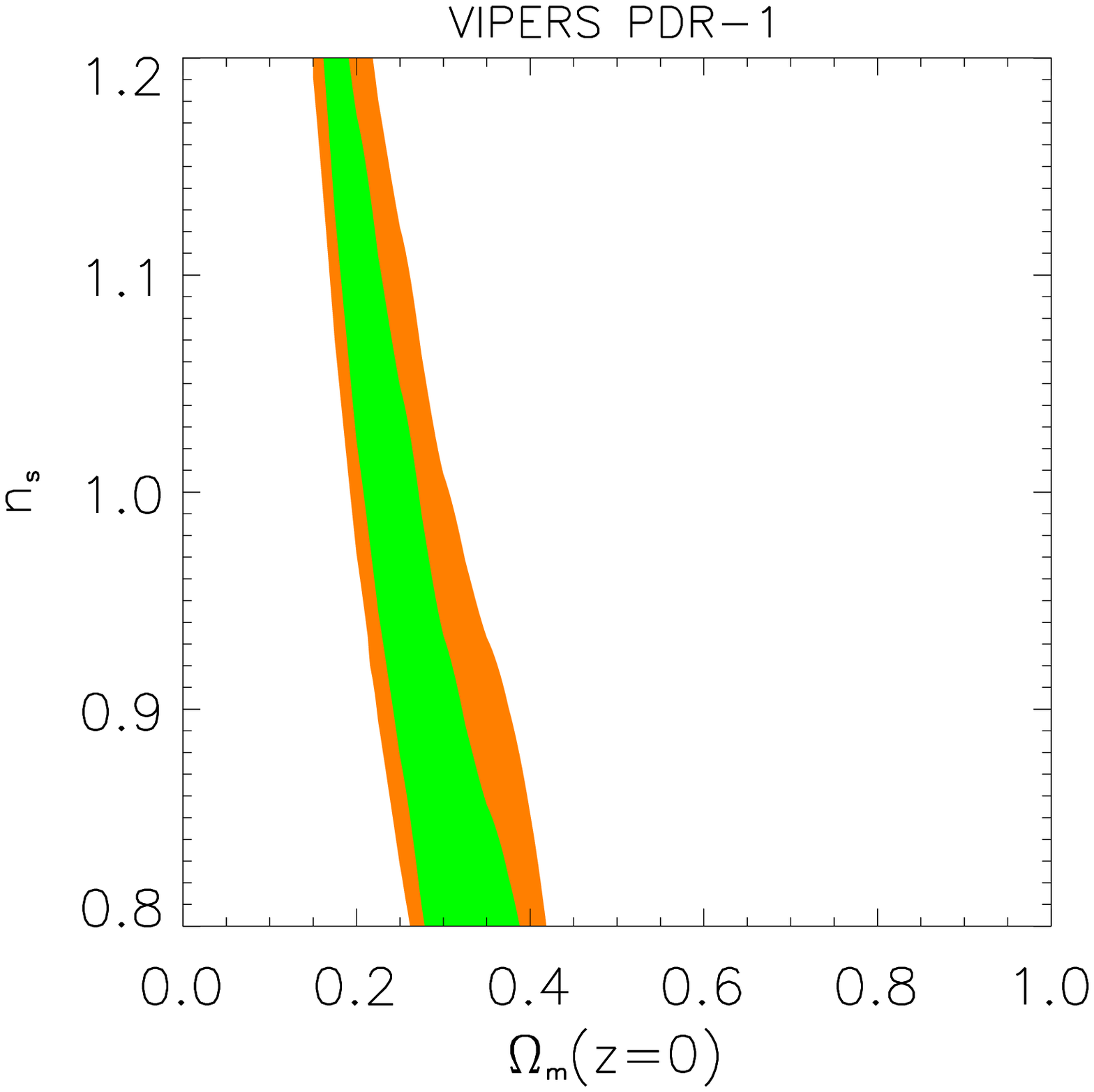}}
  \caption{\small {\it Left: } Two-dimensional marginalized
    constraints on $\Omega_{m,0}$ and $H_0$ obtained by fitting the
    $\eta$ measurements from VIPERS data with a flat $\Lambda$CDM
    model.  Contours are plotted for 
    ${\mathcal L}-{\mathcal L}_{min}< 2.3,6.17$.  In this case,
    $\Omega_b h^2$, $n_s$, $\sigma_{8}$, and $\sigma_{12}$ are fixed to 
    the value quoted in the text (i.e. they have a Dirac delta prior). 
    We also draw curves corresponding to degeneracy along the loci
    $\Omega_{m,0}h^2=const$ (dotted line) 
    and $\Omega_{m,0}h=const$ (dashed line).  {\it
      Right}: The same but in the plane [$\Omega_{m,0}$,
    $n_s$]. } 
\label{figlik2}
\end{figure*}

It is important to remark that our estimates of $\Omega_m$ using the
observed values of $\eta$ were obtained without including the full CMB
likelihood, as this would have forced a very strong prior on $\Omega_m
h^2$.  We did, however, include priors on the parameters $H_0$ and
$n_s$ on which the analysis is sensitive, as discussed earlier.  
The additional degrees of freedom arising when
these two priors in the likelihood analysis are relaxed are
shown in Fig.~\ref{figlik2}.  
At least for the range explored, the degeneracy axis in the
$(\Omega_{m},h)$ plane develops along the line $\Omega_m
h^2=const$. This is different from measurements in the linear
regime, where the degeneracy roughly follows an $\Omega_m h=const$ locus.
In principle, a redshift survey larger than (but as dense as) VIPERS
could be able to simultaneously constrain $\Omega_m$ and $h$ with a
single probe by combining measurements of $\eta$ with small and large
smoothing scale $R$.

We also note that the VIPERS value $\tilde
\eta_{g,R} \sim 0.14$ 
is only $\sim 7 \%$ smaller than the value we obtain from the $L_{gV}$
mock catalogues in a $\Lambda$CDM cosmology. 
Given that the best fitting and the simulated cosmological models are extremely close, 
this agreement independently confirms that systematic errors are
effectively taken care of.

%
\subsection{Discussion: specific sensitivity of $\eta$ to cosmological parameters}
\label{sec:ap}

\begin{figure*}[!htbp]
\includegraphics[scale=0.55]{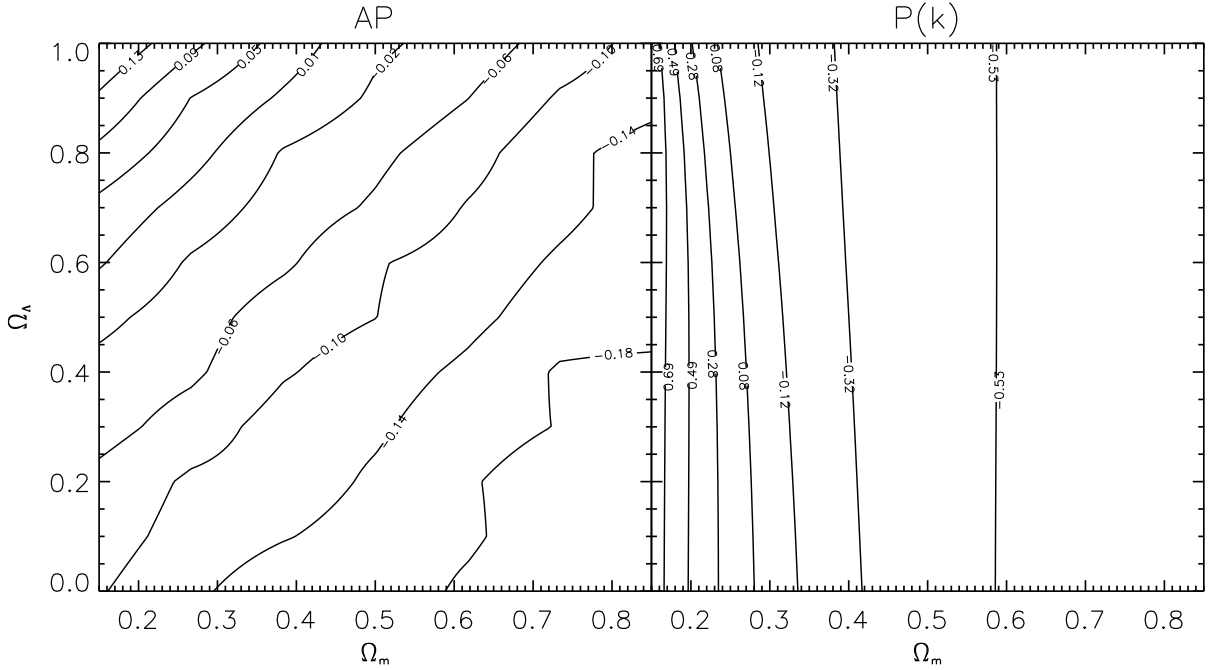}
\caption{\label{fig:appk} Sensitivity of the clustering ratio to the
  cosmological model, through {\it Left:} Isocontours of the function
  $\tilde \eta_{g,R}/\tilde \eta_{g,R}^{true}-1$ displaying the
  relative variation in the {\it measured} galaxy clustering ratio (with respect to
  the true cosmological model of the simulated data), which are induced by 
  choosing a wrong cosmology in the redshift-distance  conversion (i.e. a wrong
  guess of the cosmological parameters $[\Omega_m, \Omega_X]$).  The
  redshift-space galaxy clustering ratio (left-hand side of Eq.~\ref{coreznl})
  is measured on scales $(R,r)=(5,15)h^{-1}$Mpc from the
  stacking of the 31 $L_h$ halo catalogues (which were built in a
  cosmology with $\Omega_m=0.27, \Omega_{\Lambda}=0.73$). 
  The isocontours measure the strength of the Alcock-Paczynski geometric
  distortions, the deformation of the cell within which galaxies
  are counted.  {\it Right:} The corresponding variation of the {\it
    predicted} mass clustering ratio (right-hand side of Eq.~\ref{coreznl}) $\tilde
  \eta_{R}/\tilde \eta_{R}^{true}-1$ with varying
  cosmology.  This panel evidences the sensitivity of $\eta$ to $\Omega_m$
  through the shape of the transfer function.
  }
\end{figure*}

In Fig.~\ref{figetagvsz}, we provide an explicit example of 
how the $\eta$ test works by plotting both the 
measurements (open squares) and the expectation of the full
non-linear model (long-dashed line) when a different cosmology is
adopted. In this case, $\Omega_m$ has been set to 0.4 within the same
flat $\Lambda$ model. The choice
of a slightly wrong cosmology makes the data and the model incompatible. On one
side, the distortion of the cosmological volume of the cell due to
the calculation of distances results in a smaller measured
value of $\tilde\eta_{g,R}$. 
At the same time, the theoretical expectation is even lower due to
the different 
shape of $P(k)$ implied by the higher $\Omega_m$.  

In terms of specific cosmological parameters, the right-hand side of
Eq.~(\ref{coreznl}) depends on the shape of the matter power spectrum in real
space. This means it is directly sensitive to the primordial
index $n_s$ and to the parameters determining the size of the horizon scale at
matter-radiation equality \footnote{In our analysis, we neglect the possible effect of massive neutrinos that
would add extra parameters.}, or $\Omega_m h^2$ and $\Omega_b h^2$. Since we pushed the analysis
into the weakly non-linear regimes in this case, two additional parameters appear in the
theoretical model (\ref{coreznl}) with respect to the linear analysis
of BM13: the normalization of the power spectrum $\sigma_8$, which
enters the expression of the shape of the quasi-linear power spectrum
of \citet{Smith03} and the galaxy pairwise velocity dispersion
$\sigma_{12}$, which parameterizes the suppression of small-scale
power produced by disordered peculiar velocities. The distance-redshift relation enters the
left-hand side of Eq.~(\ref{coreznl}) changing the volume and shape of each sphere and their relative distance in the directions  parallel and perpendicular to the observer line's of sight (Alcock-Paczynski effect). 
Together, these add some sensitivity to the
density parameter, $\Omega_x$, and equation
of state, $w$, of dark energy. 

It is interesting to discuss the relative contribution of the
geometrical and power spectrum shape, which
are shown and compared in Fig.~\ref{fig:appk}. The
geometric distortions in $\eta_{g,R}(r)$ (for $R=5 h^{-1}$Mpc and
$r=3R$) are equally sensitive to both the dark matter and the dark
energy densities as indicated by the degeneracy axes
(i.e. the isocontours), which are tilted by $\sim 45^{\circ}$ in the
$[\Omega_m, \Omega_X]$ plane. The absolute amplitude of the
distance-redshift translation, however, is weak.  
The clustering ratio varies from $10$\% to $-20$\% around its
true value (estimated in the true cosmology) when the matter density parameter goes from 
$0.2$ to $0.8$ as shown in Fig.~\ref{fig:appk}.
The mass clustering ratio is extremely sensitive to
the matter density parameter $\Omega_m$. A change by $\sim 0.1(-0.1)$
of the true value (in this case $\Omega_m=0.27$) induces a change in
$\eta_R$ of $\sim -30\%(+60\%)$. Note also the extremely weak
dependence of the isocontours on the dark energy parameter.  This
arises because we parameterize the local normalization $\sigma_8$ in
terms of the seven-year WMAP normalization prior
$\Delta^2_{\mathcal{R}}(k_{w_{map}})=(2.208 \pm 0.078)\times10^{ -9}$, where
$k_{w_{map}}=0.027$ Mpc$^{-1}$ \citep{koma}. Mapping the primordial
(linear) curvature perturbations $\mathcal{R}$ into the present day
$rms$ of the matter density fluctuations, $\sigma_8$, causes the
non-linear power spectrum to acquire an extra sensitivity to the dark
energy content (and to the gravitational model) of the universe.

In conclusion, we see that the 
cosmological signal measured by $\eta$ from the VIPERS data is
primarily encoded in the shape of the power spectrum rather than in
space geometry. To obtain meaningful limits on $\Omega_v$
(or, more directly, the degree of matter domination at $z\simeq 1$),
Fig.~\ref{fig:appk} indicates that it is necessary to measure
$\Omega_m$ to a precision of better than about 5\%. This is 
approximately the precision achieved in the current analysis, so it
will be worth revisiting the geometrical aspect of the $\eta$ test when
the final VIPERS dataset is complete.

\section{Summary}

In this paper, we have applied the {\it galaxy
  clustering ratio}  proposed by \citet{bm} to a sub set
of the new VIPERS PDR-1 redshift catalogue which includes 26,611 galaxies. 
A large part of the paper has been dedicated to verifying the
robustness of applying this statistics to mildly
non-linear scales, where non-linear clustering, redshift-space
distortions, and scale-dependence of biasing need to be accounted for. 

Ideally, $\eta_{g,R}(r)$ should be estimated in the linear regime
(large $R$ and $r$ values), where its amplitude is theoretically
predicted in a straightforward way.  Nonetheless, we
have shown that a simple, theoretically motivated modelling of the
expected non-linear corrections is sufficient to preserve its value as
a powerful cosmological probe also in the mildly non-linear regime in this paper.
These corrections include: (a) The effect of random peculiar
velocities, or non-linear redshift distortions on small scales: we have also shown that
the effect of redshift errors can be incorporated within the same
scheme once an effective pairwise dispersion is considered. 
(b) The scale dependence of bias: we have shown that this effect  does not introduce a significant systematic error on the measurements
of $\eta$ for realistic
dependences.  
This is not simply a  question of assuming a mild dependence on scale
of the bias  but that its
effect can be minimized by a careful choice of the scales of smoothing ($R$)
and correlation ($r$) in terms of which $\eta$ is defined.

Based on these results, we have used $\eta$ to extract
cosmological information from the new PDR-1 data of the VIPERS survey,
 which is limited to $z>0.65$. We
first split the sample into two redshift slices with a
mean redshift $z= 0.79$ and  $z=1.07$. This exercise yielded two
independent estimates of the matter density parameter at the current 
epoch, $\Omega_{m,0}=0.281^{+0.054}_{-0.043} $ and $0.268_{-0.029}^{+0.034}$, respectively.  To obtain these estimates, we included
external priors on $\Omega_bh^2, h, n_s$ and 
$\Delta_{\mathcal{R}}^2$, but not the strong one on $\Omega_{m}
h^2$ from WMAP.  We also analysed the full redshift range $0.65<z<1.2$
as a single sample, obtaining a value $\Omega_{m,0}=0.270_{-0.025}^{+0.029}$. 

The two measurements of $\eta$ from VIPERS with
that from the SDSS by BM13 provide us with three 
independent estimates at three different cosmic epochs based on 
the same external priors.  These allow us to: {\it i)} confirm the 
consistency of the standard $\Lambda$CDM cosmological model over a
wide redshift baseline, covering almost half the age of the Universe, and {\it ii)} 
obtain a stronger constraint on the local value of $\Omega_m$ than
allowed by the two redshift surveys alone ($\sim 7\% $), corresponding
to a transition redshift between the matter and cosmological-constant
dominated epochs of $z_{md}=0.395 \pm 0.04$.

It is interesting to compare these figures to those we previously
obtained using the 
angular power spectrum of the full CFHTLS parent galaxy sample (sliced using the same
VIPERS colour selection criteria), which are coupled to an earlier
VIPERS catalogue \citep{Granett12}.  This sample, assuming flatness
and letting the matter density as a free fitting parameter (with Dirac priors on all other relevant cosmological quantities), yielded $\Omega_{m,0}=0.30 \pm 0.06$. 

The VIPERS measurement of $\Omega_{m, 0} $
is also consistent with the estimate obtained by BM13
using the same technique applied to the SDSS LRG sample at $z\sim
0.3$, $\Omega_{m,0}=0.275\pm 0.020$.  
The combination of these two estimates of $\Omega_{m,0}$, as obtained
from data that cover the large range $0 < z < 1.2$, gives a best
value of $\Omega_{m,0}=0.274\pm 0.017$.  For comparison, the recent combination
of the Planck TT power spectrum with WMAP polarization gives
$\Omega_{m,0}=0.315\pm 0.017$. Note that the apparent tension between those measurements
is mainly due to the prior on $H_0$ coming from the HST measurements. Indeed, a lower value of $H_0$ would result in a larger matter density.

\begin{acknowledgements}

JB and CM acknowledge useful discussions with F. Bernardeau,
E. Gazta\~naga, and I. Szapudi. We thank the referee for well-directed comments. We acknowledge the crucial
contribution of the ESO staff for the management of service
observations. In particular, we are deeply grateful to M. Hilker for
his constant help and support of this programme. Italian participation
to VIPERS has been funded by INAF through PRIN 2008 and 2010
programmes. LG and BJG acknowledges support of the European Research
Council through the Darklight ERC Advanced Research Grant (\#
291521). OLF acknowledges support of the European Research Council
through the EARLY ERC Advanced Research Grant (\# 268107). Polish
participants have been supported by the Polish Ministry of Science
(grant N N203 51 29 38), the Polish-Swiss Astro Project (co-financed
by a grant from Switzerland, through the Swiss Contribution to the
enlarged European Union), the European Associated Laboratory
Astrophysics Poland-France HECOLS and a Japan Society for the
Promotion of Science (JSPS) Postdoctoral Fellowship for Foreign
Researchers (P11802). GDL acknowledges financial support from the
European Research Council under the European Community's Seventh
Framework Programme (FP7/2007-2013)/ERC grant agreement n. 202781. WJP
and RT acknowledge financial support from the European Research
Council under the European Community's Seventh Framework Programme
(FP7/2007-2013)/ERC grant agreement n. 202686. WJP is also grateful
for support from the UK Science and Technology Facilities Council
through the grant ST/I001204/1. EB, FM and LM acknowledge the support
from grants ASI-INAF I/023/12/0 and PRIN MIUR 2010-2011. CM is
grateful for support from specific project funding of the {\it
Institut Universitaire de France} and the LABEX OCEVU.

\end{acknowledgements}

\end{document}